\documentclass{jfm}

\usepackage{siunitx}
\usepackage[font=small,labelfont=bf,justification=justified]{caption}
\usepackage{wrapfig}
\usepackage{graphicx}
\usepackage{newtxtext}
\usepackage{newtxmath}
\usepackage{natbib}
\usepackage[export]{adjustbox}
\usepackage{hyperref}
\hypersetup{
    colorlinks = true,
    urlcolor   = blue,
    citecolor  = black,
}
\usepackage{soul}
\usepackage{xcolor}
\usepackage{caption}
\usepackage{float}
\usepackage[labelsep=period]{caption}
\usepackage{textcomp}
\usepackage{filecontents}
\usepackage{fixltx2e}

\usepackage{hyperref}
\usepackage{amsmath}
\usepackage{epstopdf}
\usepackage{graphicx}
\usepackage{color}
\usepackage{soul}
\usepackage{setspace} 
\usepackage{amssymb}
\usepackage{amsmath}
\usepackage{multicol}
\usepackage{latexsym}
\usepackage{subfigure}
\usepackage{multirow}
\usepackage{setspace}

\usepackage{multirow}
\usepackage{arydshln}
\usepackage{amsmath,amssymb}
\usepackage{ragged2e}
\usepackage{tikz}
\usetikzlibrary{calc,topaths}
\usepackage{notoccite}

\usepackage{xcolor}
\usepackage{setspace} 
\usepackage{textcomp}

\newcommand{\RomanNumeralCaps}[1]
\linenumbers
\renewcommand{\figurename}{Figure}

\title{\justify{Three-dimensional clustering characteristics of large-stokes number sprays interacting with turbulent swirling co-flows}}

\author{Ali Rostami, Ri Li,
 \and Sina Kheirkhah\corresp{\email{sina.kheirkhah@ubc.ca}}}

\affiliation{\aff{}School of Engineering, University of British Columbia, Kelowna, British Columbia, Canada, V1V1V7}

\begin{document}
\maketitle
 
\begin{abstract}
Three-dimensional (3D) clustering characteristics of large-stokes number sprays interacting with turbulent swirling co-flows are investigated experimentally. The Astigmatic Interferometric Particle Imaging (AIPI) technique is utilized for simultaneous measurement of the spray droplets position in 3D space and their corresponding diameter. The Stokes number estimated based on the Kolmogorov time scale varies from 34 to 142. The results show that the degree of droplet clustering plateaus at about 0.4 and at large Stokes numbers. It is obtained that the mean length scale of the clusters normalized by the Kolmogorov length scale follows a power-law relation with the Stokes number, and the mean void length scale normalized by the integral length scale plateaus at about 1.5. It is shown that the ratio of the number density of the droplets residing within the clusters to the global number density increases with increasing the Stokes number and is about 8 for the largest stokes number examined in this study. The joint characteristics of cluster's normalized volume and the mean diameter of droplets residing within the clusters show that small-volume clusters accommodate droplets with a relatively broad range of diameter. However, large clusters carry droplets with the most probable diameter. The developed AIPI technique in the present study and the corresponding spray characteristics are of importance for engineering applications that aim to understand the 3D clustering characteristics of large-stokes number droplets sprayed into turbulent swirling co-flows.
\end{abstract}
\begin{keywords} Particle-laden flows,  Turbulent swirling flows, AIPI, Sprays
\end{keywords}

\section{Introduction}
\label{sec:Introduction}

Understanding the interaction of turbulent flows and sprays is key for the design and development of many engineering applications, for example, aviation gas turbine engine combustors, liquid-fueled industrial burners, industrial driers, and cooling towers, see~\cite{jenny2012modeling, sommerfeld1998experimental, Crowe2011Multiphase}. Of importance for developing the above understanding is simultaneously measuring the spray droplets position in three-dimensional (3D) space along with the droplets' diameter. Despite the importance of the 3D measurements, the majority of past experimental investigations used techniques for measurements in one-dimensional or two-dimensional (2D) spaces, see for example ~\cite{allen1995planar,kourmatzis2015characterization,vignat2021investigation,skeen2015simultaneous}. As such, experimentally measured location and diameter of the spray droplets in 3D space is rarely available. The present study is motivated by developing the above knowledge for non-reacting sprays injected in swirling flows, with relevance to gas turbine engine combustors.

The 2D (planar) measurements of past studies concerning sprays injected in turbulent flows showed that the droplets are not randomly distributed in the measurement plane, and they form clusters, see~\cite{manish2021optical}. For the 2D measurements, the characteristics of clusters have been studied for both sprays injected in turbulent flows as well as particle-laden flows; and, many review papers have been published, see for example,~\cite{poelma2006particle, balachandar2010turbulent}. In the past studies, a laser sheet was employed to illuminate the droplets in a plane, the Mie scattered light from the droplets/particles was collected, the light intensity was used to obtain the location of the droplets/particles centres, these centres were used to identify the Vorono\"{i} cells~\cite{monchaux2010preferential}, and finally, the Probability Density Function (PDF) of the Vorno\"{i} cells area was calculated and compared against the PDF of the Vorono\"{i} cells area provided the droplets/particles were randomly distributed in the 2D space, see \cite{frankel2016settling, tagawa2012three, boddapati2020novel}. Such comparison of the PDFs showed that there existed two Vorono\"{i} cells area below and above which the PDFs of the experimentally obtained Vorono\"{i} cells area were substantially larger than those of the droplets/particles provided they were randomly distributed. The cells whose areas are smaller (larger) than the smaller (larger) intersection area were separately merged provided they featured overlapping edges and were referred to as clusters (voids), as detailed in~\cite{sumbekova2017preferential}. 

Among many characteristics, the degree of the clustering, the length scale of the clusters and voids, the number densities of the droplets that reside within the clusters and voids, and the Joint PDF of the diameter of the droplets inside the clusters/voids and the size of the clusters/voids have been studied in past investigations, see for example~\cite{sumbekova2017preferential, obligado2014preferential,rostami2023separate}. The degree of clustering is defined as $(\sigma-\sigma_\mathrm{RPP})/\sigma_\mathrm{RPP}$, with $\sigma$ and $\sigma_\mathrm{RPP}$ being the root-mean-square (RMS) of the normalized Vorono\"{i} cells area estimated from the measurements and from synthetically distributed droplets/particles following a Random Poison Process (RPP), respectively. The studies of~\cite{monchaux2010preferential, obligado2014preferential,sumbekova2017preferential} showed that the degree of clustering can be primarily influenced by three non-dimensional parameters: the Stokes number ($St$) estimated based on the Kolmogorov time scale, Taylor length scale-based Reynolds number ($Re_\lambda$), and the volume fraction of the droplets ($\phi_\mathrm{v}$). Specifically, \cite{sumbekova2017preferential} showed that in turbulent particle-laden flows with relatively large values of the Taylor length scale-based Reynolds number ($Re_\lambda \gtrsim 200$), though the Stokes number did not substantially influence the degree of clustering, this parameter followed a power-law relation with $Re_\lambda$ and $\phi_\mathrm{v}$. The exponents of the power-law relation between the degree of clustering and $Re_\lambda$ and $\phi_\mathrm{v}$ were 1 and 0.5, respectively, as reported in~\cite{sumbekova2017preferential}. Compared to flows with relatively large $Re_\lambda$, for moderately turbulent particle-laden flows ($Re_\lambda \lesssim 200$), the studies of~\cite{monchaux2010preferential, obligado2014preferential} showed that the degree of clustering featured a peak near $St\approx 2-4$ for $St\lesssim 10$. Motivated by understanding the clustering characteristics at Stokes numbers that are closer to those of gas turbine engine combustors, \cite{rostami2023separate} studied sprays injected in turbulent co-flows. In \cite{rostami2023separate}, it was shown that the degree of clustering plateaus at about 0.3 for flows with Stokes numbers as large as about 25.

The clusters ($L_\mathrm{c}$) and voids ($L_\mathrm{v}$) length scales were defined as the square root of the corresponding areas in~\cite{aliseda2002effect}. The studies of \cite{sumbekova2017preferential} showed that both of these length scales were influenced by $Re_\lambda$ and $St$. Generally, $L_\mathrm{c}$ was about one to two orders of magnitude larger than the Kolmogorov length scale and increasing both Stokes and the turbulent Reynolds number increased the cluster length scale normalized by the Kolmogorov length scale. The study of \cite{rostami2023separate} showed that $L_\mathrm{v}$ was on the order of the integral length scale ($\Lambda$), and $L_\mathrm{v}/\Lambda$ did not change significantly by varying $Re_\lambda$ or $St$. 

The joint characteristics of droplets and clusters/voids for sprays injected in turbulent co-flows were studied in the authors' previous work, see~\cite{rostami2023separate}. It was shown that, for $St \lesssim 10$, increasing this parameter reduced the number density of the droplets that reside within both clusters and voids. However, for $10 \lesssim St \lesssim 25$, the number density of the droplets inside clusters and voids were about 0.45 and 0.06 mm$^{-2}$, respectively. For all background turbulent flow conditions tested in \cite{rostami2023separate}, the number densities of droplets that resided inside the clusters and voids were about 5.5 and 0.8 times the global number density, respectively. Though trends from the joint PDF of the diameter of the droplets that resided within voids and voids normalized area could not be obtained in~\cite{rostami2023separate}, conclusions were drawn from the joint PDF of the droplet diameter within the cluster and the cluster normalized area. It was concluded in \cite{rostami2023separate} that the  majority of the droplets reside within clusters with areas smaller than the average cluster area, and a wide range of clusters accommodate droplets with the most probable diameter.   

As demonstrated in~\cite{rostami2023separate}, key to understanding the joint characteristics of sprays and clusters is the acquisition of the Mie scattered light of the droplets along with the interference pattern as a result of the reflected and refracted light from the droplets. In~\cite{rostami2023separate}, the interference patterns were acquired using the Interferometric Laser Imaging for Droplet Sizing (ILIDS) technique, similar to the studies of for example~\cite{maeda2002improvements, zama2005simultaneous, hardalupas2010simultaneous, qieni2016high, boddapati2020novel}. Though the ILIDS and its combination with the Mie scattering technique are enabling, as they facilitate understanding the joint characteristics of sprays, the simultaneous information regarding the droplets diameter and their position in 3D space could not be extracted from the ILIDS technique. In fact, the results of past Direct Numerical Simulations for particles interacting with turbulence in a box, see for example~\cite{onishi2014collision}, suggested that the clustering characteristics obtained from 2D data were different from those obtained from 3D data. Aiming to measure the diameter of the droplets in 3D space,~\cite{ouldarbi2015simultaneous, shen2012cylindrical} made improvements to the ILIDS hardware, with the improved version of the ILIDS technique referred to as Astigmatic Interferometric Particle Imaging (AIPI), see~\cite{wu2022accurate,wen2021characterization}.

The above background allows for identifying two gaps in the literature. First, though past investigations that studied the clustering characteristics are of significant importance, the largest tested Stokes number was about 25, which is at least several folds smaller than those relevant to some engineering applications, such as gas turbine combustors. Second, although the AIPI technique is recently developed and allows for obtaining both the 3D position and the diameter of the droplets, no investigation has implemented this technique to characterize the clustering of the droplets in turbulent flows. The objective of the present study is to develop our knowledge concerning the joint characteristics of spray droplet diameter and clusters using 3D measurements and for relatively large Stokes numbers.

\section{Experimental methodology}\label{sec:Experimental methodology}
The details of the experimental setup, diagnostics, coordinate systems, and test conditions are presented in this section. 

\subsection{Experimental setup}\label{subsec:Setup}
The experimental setup included air and water delivery systems as well as a flow apparatus. The air delivery system included an Atlas Copco compressor (model GA37FF), and the air flow rate was measured using an MCRH 5000 ALICAT mass flow controller. The water delivery system included a nitrogen bottle to purge water from a pressurized tank as well as a PCD-100-PSIG dual valve pressure controller from ALICAT. For all test conditions, the reservoir pressure was fixed at 62.1 kPa gauge. The liquid flow rate was calibrated following the procedure detailed in \cite{rostami2023separate}. It was measured that the above reservoir pressure corresponded to an injection of 13.8 grams of water per minute. Both the air and water delivery systems were identical to those used in our earlier study, see \cite{rostami2023separate}. 

\begin{figure}
	\centerline{\includegraphics[width=0.9\textwidth]{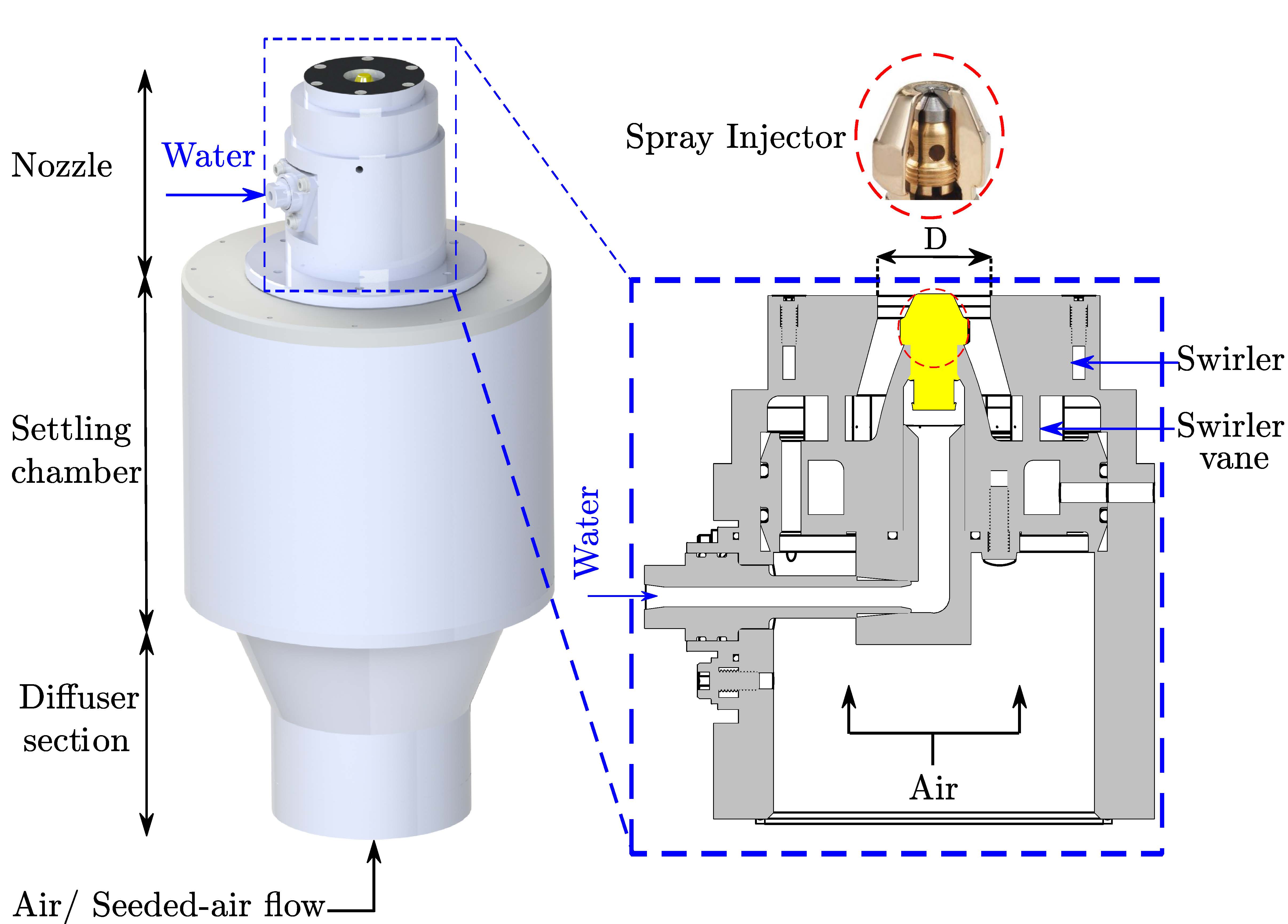}}
	\caption{The flow apparatus, which is composed of a diffuser section, a settling chamber, and a nozzle. The inset on the right-hand-side is the technical drawing of the nozzle cross-section as well as a picture of the atomizer tip.}
	\label{Fig:Setup}
\end{figure}

The flow apparatus included a diffuser section, a setting chamber, and a nozzle, which are shown in Fig.~\ref{Fig:Setup}. The area ratio of the diffuser was about four, and the settling chamber was equipped with five equally spaced mesh screens, identical to \cite{kheirkhah2016experimental,kheirkhah2013turbulent,mohammadnejad2022new,rostami2023separate}. Compared to these studies, the nozzle section was different and was similar to~\cite{Krebbers2023113016}. Specifically, the nozzle section included a Delavan pressure swirl atomizer (model S0075-60S1) and a swirler, with details shown in the inset of Fig.~\ref{Fig:Setup}. In the past studies, e.g.~\cite{wang2019soot}, the combination of the nozzle and a combustion chamber was referred to as the Gas Turbine Model Combustor. Compared to \cite{Krebbers2023113016,wang2019soot}, improving the quality of our measurements and accommodating a relatively large Field-Of-View (FOV), the combustion chamber was removed. Also compared to past studies, water (instead of Jet A-1) was the working fluid and the flow was non-reacting in the present study.

\subsection{Diagnostics}\label{subsec:Diagnostics}
\begin{figure}
	\centerline{\includegraphics[width=0.95\textwidth]{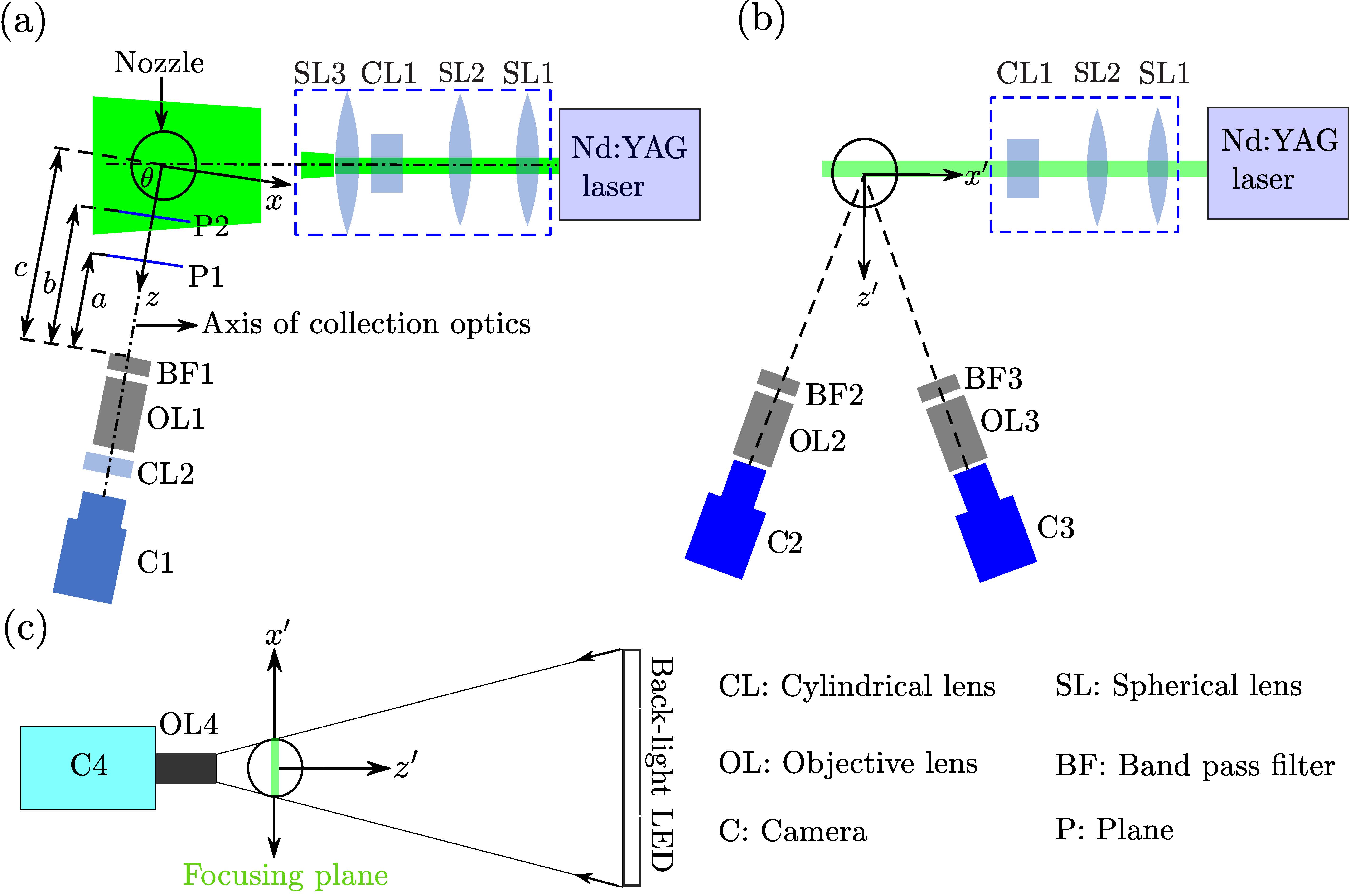}}
	\caption{The utilized diagnostics. (a)--(c) are the schematic arrangement of Astigmatic Interferometric Particle Imaging, Stereoscopic Particle Image Velocimetry, and shadowgraphy technique, respectively. The dimensions are not to scale. }
	\label{Fig:Diagnostics}
\end{figure}

Three diagnostics, namely, Astigmatic Interferometric Particle Imaging, Stereoscopic Particle Image Velocimetry (SPIV), and shadowgraphy were separately used in the present study. The schematic arrangement of these diagnostics are presented in Figs.~\ref{Fig:Diagnostics}(a--c), respectively. The dimensions in the figure are not to scale. The details of the utilized diagnostics are discussed in the following.

\subsubsection{Astigmatic Interferometric Particle Imaging}\label{subsubsec:AIPI}

The AIPI technique was used for the simultaneous measurement of the 3D location of spray droplets and their corresponding diameter. The hardware of the AIPI technique consisted of an Nd:YAG dual cavity pulsed laser (model: Evergreen PIV 200), volume illumination optics, an Andor’s Zyla 5.5 sCMOS camera (C1), and collection optics for C1. The laser produced a 532~nm laser beam, with a beam diameter of about 5~mm. For all experiments, the laser was operated at 45\% of its maximum energy to avoid saturation of the AIPI images. The laser beam was converted into a laser slab using the volume illumination optics (see the blue-dashed box in Fig.~\ref{Fig:Diagnostics}(a)). The volume illumination optics included the LaVision divergent sheet optics, which were two spherical lenses (SL1 and SL2), a cylindrical lens (CL1)  with a focal length ($f$) of $-20$~mm, and a spherical lens (SL3 from Thorlabs, model LA1145-AB-ML) with $f = 75$~mm. It is important to note that the combination of SL1, SL2, and CL1 allowed for generating a rectangular laser slab, with a thickness of about 10~mm. However, such thickness was not sufficient for the purposes of the volumetric illumination of the spray in the present study, and as such, SL3 was additionally installed. This last spherical lens allowed for increasing the thickness of the illuminated volume from about 10 to 60~mm near the nozzle centreline, which was 1060~mm distant from the centreline of SL3. For presentation purposes, this distance is shortened in Fig.~\ref{Fig:Diagnostics}(a). The collection optics included a 532$\pm$10~nm bandpass filter (BF1), a Macro Sigma objective lens (OL1, with $f = 105$~mm), and a cylindrical lens (CL2, with $f = 250$~mm). CL2 was installed on a rotary mount (model CLR1 from Thorlabs) which allowed to rotate CL2 with respect to the axis of the collection optics, see Fig.~\ref{Fig:Diagnostics}(a). CL2 was positioned between the camera sensor and OL1 to create astigmatic effect, similar to the studies of~\cite{wen2021characterization}. The axis of the collection optics was positioned such that it made an angle of $\theta = 70^\mathrm{o}$ with the laser beam axis, see the angle between the dotted-dashed lines in Fig.~\ref{Fig:Diagnostics}(a). This angle was selected as it yielded the best clarity of interferometric images, following~\cite{sahu2011experimental}. For each test condition (discussed later), 2000 AIPI images were acquired with the acquisition frequency of 10~Hz. 

\begin{figure}
 	\centerline{\includegraphics[width=0.55\textwidth]{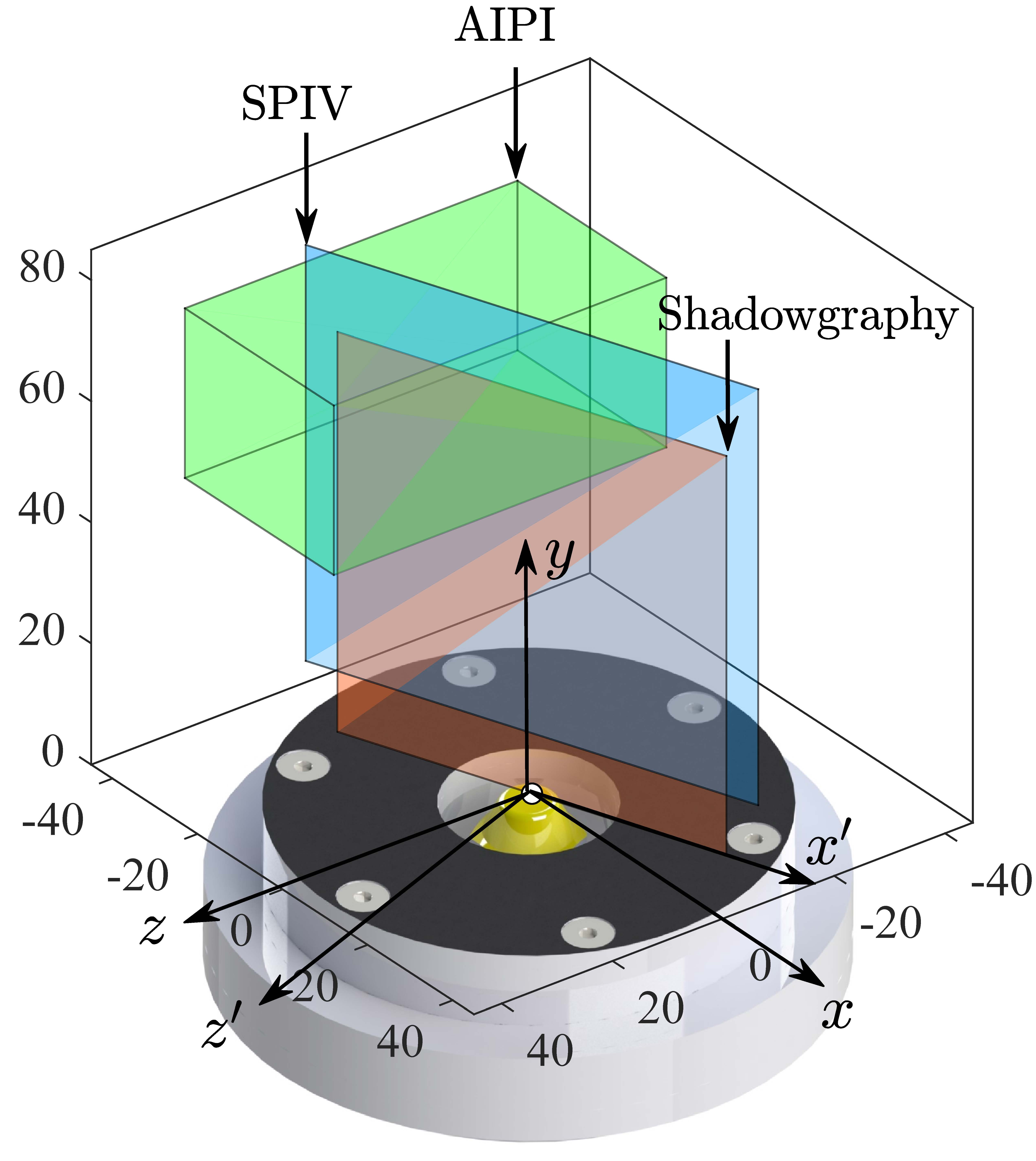}}
 	\caption{The green box, blue slab, and red slab present the AIPI, SPIV, and shadowgraphy regions-of-interest, respectively. $(x,y,z)$ is the coordinate system used for AIPI, and $(x',y,z')$ is that used for the SPIV and the shadowgraphy techniques.} 	
 	\label{Fig:Coordinate system}
 \end{figure}

A Cartesian coordinate system was used to locate the AIPI region-of-interest. This coordinate system is shown by $x$, $y$, and $z$ axes in Fig.~\ref{Fig:Coordinate system}. The origin of this coordinate system coincides with the intersection of the spray injector centreline ($y-$axis) and the exit plane of the nozzle. The $z-$axis coincides with the axis of the collection optics and points towards the camera sensor. $x-$axis is perpendicular to both $y$ and $z$-axes. The AIPI region-of-interest is shown by the transparent green box in Fig.~\ref{Fig:Coordinate system}. This region extends from -42.5 to -7.5, 43.2 to 71.2, and -30.0 to 30.0~mm, along the $x-$, $y-$, and $z-$axes, respectively. Thus, the AIPI volume-of-interest ($V_\mathrm{AIPI}$) is $35\times28\times60=58800~\mathrm{mm}^3$.

\subsubsection{Stereoscopic Particle Image Velocimetry}
The SPIV experiments were performed to characterize the background turbulent swirling air flow. The SPIV hardware included a laser (identical to that of the AIPI technique), optics to generate a laser sheet, a programmable timing unit (model PTU~X from Lavision), and two Imager CX5 sCMOS cameras (see C2 and C3 in Fig.~\ref{Fig:Diagnostics}(b)) which were equipped with collection optics. The collection optics included two Scheimpflug adapters, SAMYANG lenses with $f = 85$~mm (see OL2 and OL3 in Fig.~\ref{Fig:Diagnostics}(b)), and two 532$\pm$10~nm bandpass filters (see BF2 and BF3 in Fig.~\ref{Fig:Diagnostics}(b)). The SPIV cameras were placed in the backward-forward scattering arrangement, and the angle between the viewing direction of the cameras (the angle between two black dashed lines in Fig.~\ref{Fig:Diagnostics}(b)) was set to $21^{\circ}$. For each SPIV experiment, 1200 image pairs were acquired at an acquisition frequency of 10~Hz.

The SPIV experiments were performed while no spray was injected into the air flow. A flow seeder (model PB110 from LaVision) was used to add Aerosil{\textregistered} particles into the air flow. The most probable single-particle diameter of the Aerosil{\textregistered} particles was about 12~nm, which was provided by the producer. The flow seeder used a magnetic stirrer to avoid agglomeration of the particles. Using the formulation provided in \cite{jacobs2007iced}, the Stokes number of agglomerated seed particles of 1~$\mu$m was less than 0.008, rendering the utilized seed particles acceptable for the SPIV measurements.

The FOV of the SPIV experiments is shown in Fig.~\ref{Fig:Coordinate system}, using the blue colour. The direction of the laser beam for the SPIV experiments was identical to that of the AIPI (along negative $x'$) as shown in Figs.~\ref{Fig:Diagnostics}(a and b). As a result, the FOV of the SPIV is tilted by $90^\mathrm{o}-\theta$, which is $20^\mathrm{o}$ with respect to the $x-y$ plane. For the SPIV experiments, a Cartesian coordinate system was used, which is shown by $x'-$, $y-$, and $z'-$axes in Fig.~\ref{Fig:Coordinate system}. It is important to note that the swirling flow is axisymmetric, and as such, the velocity statistics are similar in all azimuthal planes that coincide with the $y-$axis. The extent of the SPIV FOV varied from -38.4 to 38.4 and 10.0 to 78.7~mm along the $x'-$ and $y-$axes, respectively. The above FOV corresponds to a pixel resolution of $76.8~\mathrm{mm}/2240~\mathrm{pixels}=34.3~\mu\mathrm{m}/\mathrm{pixel}$. The LaVision DaVis 10.2.1 software was used to calculate the velocity fields from the Mie scattering images of the seed particles. In the calculations, a spatial cross-correlation algorithm was used to initially calculate the velocity vectors for $128\times128~\mathrm{pixels}^{2}$ interrogation windows. This was reduced to a final interrogation window size of $32\times32~\mathrm{pixels}^{2}$ with an overlap of $50\%$ between the windows. The time separation between the two laser pulses was varied for each experiment and was calculated by dividing 25\% of the final interrogation window size and the mean bulk flow velocity. The above time separation varied from 8.2--77.0~$\mu$s, depending on the test condition. 
 
\subsubsection{Shadowgraphy}\label{subsubsec:Shadowgraphy}
The shadowgraphy technique was used to understand the effects of swirling flow on the spray formation and large droplets break-up. The shadowgraphy hardware included a back-light LED lamp (model NL-360ARC from Neewer) and a Photron Fastcam Nova S12 high-speed camera (C4 in Fig.~\ref{Fig:Diagnostics}(c)) equipped with a Macro Sigma lens, with $f = 105$~mm and its aperture size was set to $f/2.8$. The shadowgraphy images were captured for a duration of 1~s with the acquisition frequency of 10~kHz. The camera exposure time was set to 20~$\mu$s. The FOV of the shadowgraphy imaging was 64~mm along the $x'$-axis and 64~mm along the $y$ axis. The depth-of-field (DOF) for the shadowgraphy camera was calculated using the formulation from~\cite{gross2008handbook}, which is given by

\begin{equation}
\label{Eq:DOF}
DOF = \frac{2of^2N\tilde{c}(o-f)}{f^4 - \tilde{c}^2N^2(o-f)^2}.
\end{equation}
In Eq.~(\ref{Eq:DOF}), $\tilde{c}$ is the camera sensor size, $N$ is the aperture number, and $o$ is the distance between the camera sensor and the plane of focus. In the present study, $\tilde{c}$ and $o$ were 20~$\mu$m and 500~mm, respectively. Substituting the above values in Eq.~(\ref{Eq:DOF}), the depth-of-field was calculated and was about 2~mm. The shadowgraphy FOV is shown by the red slab in Fig.~\ref{Fig:Coordinate system}.

\subsection{Test conditions}\label{subsec:Testconditions}
In total, 8 test conditions were examined with details presented in Table~\ref{tab:tested onditions}. For all test conditions, the liquid flow rate was kept constant. The mean bulk flow velocity of the air flow ($U_\mathrm{b}$) was estimated by dividing the set air flow rate to the exit area of the nozzle section, which is the area of a circle with diameter $D=28.0~\mathrm{mm}$, subtracted by the surface area of the Delavan injector at its exit, see the inset of Fig.~\ref{Fig:Setup}. In the present study, $U_\mathrm{b}$ was varied from 6.1 to 27.4~$\mathrm{ms^{-1}}$, with corresponding values tabulated in the first column of Table~\ref{tab:tested onditions}. It is important to note that, for $U_\mathrm{b} < 6.1~\mathrm{ms^{-1}}$, the spray was too dense and the AIPI was not performed for these velocities. However, qualitative analysis of the spray was possible using the shadowgraphy technique, and as a result, in addition to the bulk velocities presented in Table~\ref{tab:tested onditions}, shadowgraphy experiments were performed for $U_\mathrm{b}= 0~\mathrm{ms^{-1}}$ (which corresponds to no background air flow). The Reynolds number estimated based on the mean bulk flow velocity was calculated using $Re_D=U_\mathrm{b}D/\nu$, with $\nu$ being the kinematic viscosity of the air at the laboratory temperature. The values of $Re_D$ are tabulated in the second column of Table~\ref{tab:tested onditions}.

The background turbulent flow characteristics were estimated at the intersection of the SPIV FOV and the AIPI region-of-interest. This intersection region extends from -39.9 to -7.0 and 43.2 to 71.2~mm, along the $x'$ and $y-$ axes, respectively. The root-mean-square of the streamwise (along $y-$axis) velocity fluctuations was averaged inside the above region, referred to as $\tilde{u}_\mathrm{RMS}$, and listed in the third column of Table~\ref{tab:tested onditions}. The integral length scale ($\Lambda$) was calculated using the auto-correlation of the streamwise velocity fluctuations, following the formulations presented in \cite{kheirkhah2015consumption,kheirkhah2016experimental}. Similar to these studies, the calculation of the integral length scale at spatial locations corresponding to $y \gtrsim 60.0$~mm was not possible since the auto-correlation of the velocity data did not attain a zero value. As such, the integral length scale was calculated in the intersection region of SPIV and AIPI and for $ 43.2~\mathrm{mm} \lesssim y \lesssim 60.0~\mathrm{mm}$. The values of $\Lambda$ are listed in the fourth column of Table~\ref{tab:tested onditions}. The Taylor ($\lambda$) and Kolmogorov ($\eta$) length scales were calculated using $\lambda = \Lambda(\tilde{u}_\mathrm{RMS} \Lambda/\nu)^{-0.5}$ and $\eta = \Lambda(\tilde{u}_\mathrm{RMS} \Lambda/\nu)^{-0.75}$. The values of $\lambda$ and $\eta$ are presented in the fifth and sixth columns of Table~\ref{tab:tested onditions}, respectively. For all test conditions, $\tilde{u}_\mathrm{RMS}$ varied from about 1.0 to 4.4~m/s, while the integral length scale was nearly constant ($\Lambda = 7.6-7.8~\mathrm{mm}$). The Taylor and the Kolmogorov length scales varied from about 160 to 347 and from 23 to 73~$\mu$m, respectively.

\begin{table}
	\begin{center}
		\def~{\hphantom{0}}
		\scalebox{1.0}{
			\begin{tabular}{c c c c c c c c c c c}
				
				 $U_\mathrm{b}$ ($\mathrm{ms^{-1}}$) & $Re_D$ &$\tilde{u}_\mathrm{RMS}$ ($\mathrm{ms^{-1}}$)& $\Lambda$ (mm)& $\lambda$ $(\mu \mathrm{m})$& $\eta$ $(\mu \mathrm{m})$& $SMD$ $(\mu \mathrm{m})$& $\overline{d}$ $(\mu \mathrm{m})$& $Re_\lambda$ &  $St$ & $\phi_\mathrm{v} (\times \mathrm{10^6})$ \\
				 6.1  & 11345 & 1.0  & 7.8 & 347 & 73 & 95 & 57 & 22.4 & 34  & 0.74                        \\
				 9.1  & 17018 & 1.4  & 7.8 & 285 & 54 & 85 & 52 & 27.4 & 50  & 0.53                            \\
				 12.2 & 22690 & 1.9  & 7.7 & 243 & 43 & 86 & 55 & 31.6 & 92  & 0.55                         \\
				 15.2 & 28363 & 2.4  & 7.8 & 219 & 37 & 78 & 47 & 35.5 & 89  & 0.30 \\
				 18.2 & 34036 & 3.0  & 7.6 & 196 & 31 & 75 & 46 & 38.8 & 117 & 0.34 \\
				 21.2 & 39708 & 3.5  & 7.8 & 183 & 28 & 69 & 41 & 42.5 & 117 & 0.33 \\
				 24.3 & 45381 & 3.9  & 7.8 & 174 & 26 & 70 & 39 & 44.9 & 128 & 0.35 \\ 
				 27.4 & 51054 & 4.4  & 7.6 & 160 & 23 & 67 & 37 & 47.4 & 142 & 0.35 \\

			\end{tabular}
		}
		\caption{Test conditions.}
		\label{tab:tested onditions}
	\end{center}
\end {table}
				


The formulation provided in~\cite{lefebvre2017atomization} and the AIPI data were used to estimate the Sauter Mean Diameter ($SMD$) of the spray droplets. The values of $SMD$ are listed in the seventh column of the table. The AIPI data was also used to estimate the mean droplet diameter ($\overline{d}$) which is listed in the eighth column of the table. For the conditions tested in the present study, $SMD$ and $\overline{d}$ varied from 67 to 95 and 37 to 57~$\mu$m, respectively. It should be noted that, since the calculation of $SMD$ involves the volume to surface area ratio of the droplets, it is anticipated that the Sauter Mean Diameter to be skewed toward the diameter of large droplets, resulting in the larger values of $SMD$ compared to $\overline{d}$. In summary, comparison of the results presented in the sixth to eighth columns of Table~\ref{tab:tested onditions} shows that the size of the spray droplets are on the order of the Kolmogorov length scale tested in the present investigation.

Of prime importance in studying particle-laden flows are the Taylor length scale-based Reynolds number, Stokes number, and the spray volume fraction ($\phi_\mathrm{v}$). The Taylor length scale-based Reynolds number was estimated using $Re_\lambda = \tilde{u}_\mathrm{RMS} \lambda/\nu$, with the corresponding values listed in the ninth column of Table~\ref{tab:tested onditions}. In the present study, $Re_\lambda$ varies from about 22 to 47, which are relatively moderate values compared to past particle laden-flow studies, see for example \cite{sumbekova2017preferential}. Following \cite{reade2000effect}, the Stokes number was calculated using
\begin{equation}\label{Eq:St}
St=\frac{1}{18}\frac{\rho_\mathrm{W}}{\rho_\mathrm{A}}\left(\frac{d}{\eta}\right)^{2},
\end{equation}
where $\rho_\mathrm{W}$ and $\rho_\mathrm{A}$ are the water and air densities both estimated at the laboratory temperature (21$^\mathrm{o}$C) and pressure (1~atm), respectively. In Eq.~(\ref{Eq:St}), $d$ is a characteristic size for the droplet diameter. In the present study, the mean droplet diameter was used for the estimation of the Stokes number. The values of the Stokes number are tabulated in the tenth column of Table~\ref{tab:tested onditions} and vary from about 34 to 142, which are relatively large compared to those tested in the literature. It is important to note that in the present study, the mean droplet diameter is larger than the Kolmogorov length scale, and following the studies of~\cite{xu2008motion, monchaux2012analyzing}, the Kolmogorov time scale may not be appropriate for calculating the Stokes number. Specifically, the above studies suggested using the droplet diameter instead of the Kolmogorov length scale for calculating the Stokes number. Using the formulation proposed in~\cite{xu2008motion, monchaux2012analyzing} led to the Stokes number varying from about 40 to 104. Either of the calculations discussed above leads to values of the Stokes number that are substantially larger than those examined in previous studies. Nonetheless, using either of the definitions of the Stokes number yields similar conclusions. In the present study, Eq.~(\ref{Eq:St}) was used for estimation of the Stokes number, facilitating comparisons with the results of past investigations, for example those in~\cite{rostami2023separate}.

The spray volume fraction was estimated using the formulation in~\cite{elghobashi1994predicting}, which is given by
\begin{equation}
\label{eq:phiv}
\phi_\mathrm{v} = \overline{n}\frac{V_\mathrm{d}}{V_\mathrm{AIPI}}.
\end{equation}
In Eq.~(\ref{eq:phiv}), $\overline{n}$ is the mean number of droplets within the volume of the AIPI region-of-interest (see the green box in Fig.~\ref{Fig:Coordinate system}), and $V_\mathrm{d}$ is the volume occupied by an individual droplet. For each test condition, the number of droplets, $n$, was calculated using the AIPI technique and was averaged over all collected images. The volume occupied by a droplet was approximated by $V_\mathrm{d} = (\pi/6) \overline{d}^{3}$. The values of $\phi_\mathrm{v}$ are listed in the eleventh column of Table~\ref{tab:tested onditions}. In the present study, the spray volume fraction varied from $0.35\times10^{-6}$ to $0.74\times10^{-6}$, rendering the tested sprays dilute, as suggested in~\cite{elghobashi1994predicting}. 

\section{Data reduction}\label{subsec:Data reduction}
The procedure for reducing the raw AIPI data into the three-dimensional position and diameter of spray droplets are presented in the first subsection. Then, in the second subsection, the procedure for identifying the 3D clusters and voids are discussed.

\subsection{Three-dimensional positioning and sizing of droplets}
\label{subsection:3D}
\begin{figure}
\centerline{\includegraphics[width=1\textwidth]{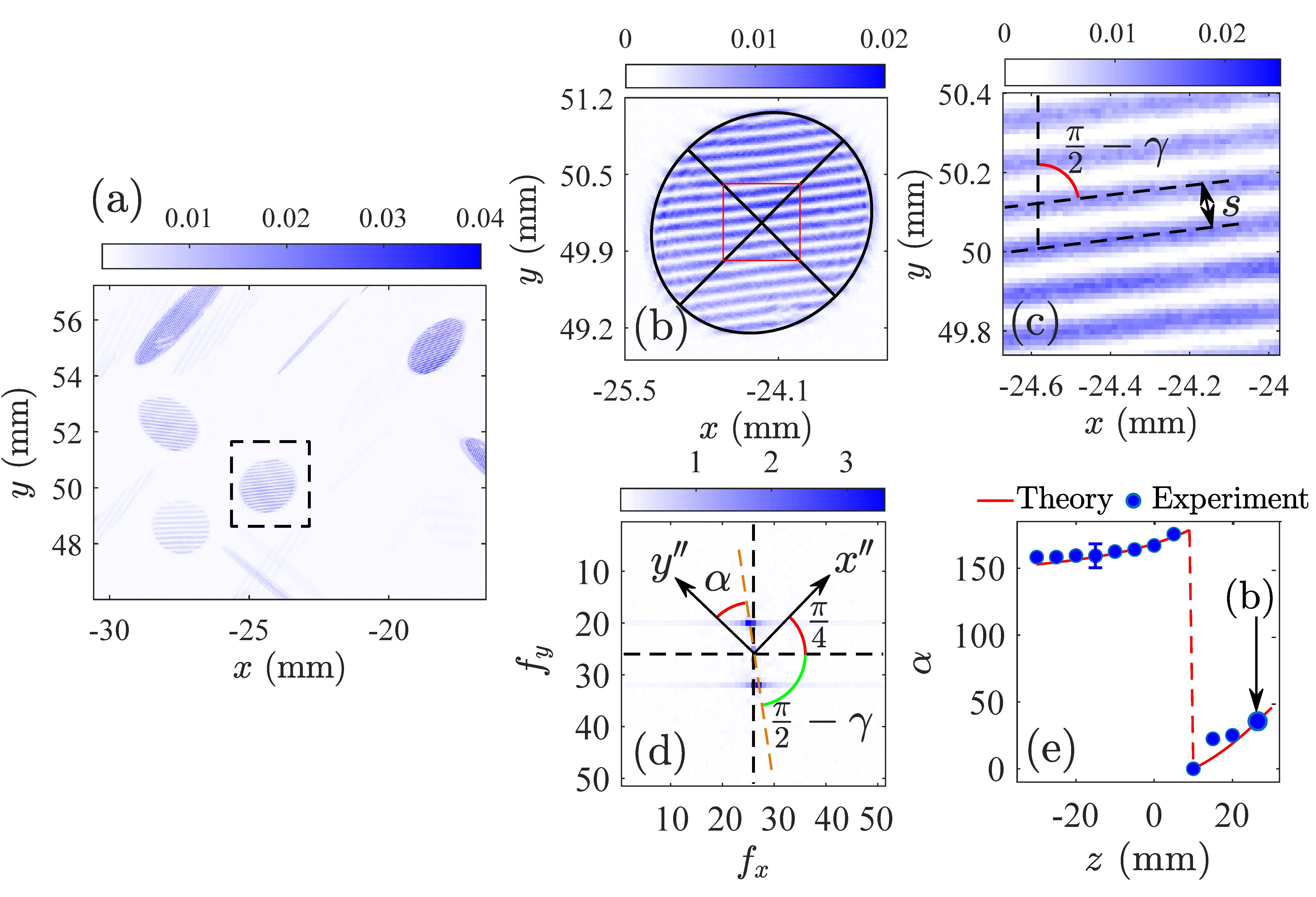}}
\caption{(a) is a cropped AIPI raw image. (b) is a sample interference pattern from (a). (c) is the inset of (b) corresponding to the red rectangle in (b). (d) is the two-dimensional Fast Fourier Transform magnitude for the interference pattern shown in (c). Overlaid on (d) are the $x''-$ and $y''-$axes which are the axes corresponding to the power and axis meridians of the cylindrical lens CL1. In panel (d), the angle between the line that connects the two peaks and $y''-$axis is $\alpha$. (e) The red solid curve presents the relation between $\alpha$ and the depth of the spray droplets obtained from Eq.~(\ref{Eq:zlocation}). In (e), the blue circular data points are obtained from the calibration experiments.}
\label{Fig:AIPI Data reduction}
\end{figure}

A cropped view of a representative image from AIPI is shown in Fig.~\ref{Fig:AIPI Data reduction}(a) which corresponds to test condition with $U_\mathrm{b}=15.2~\mathrm{ms^{-1}}$. As can bee seen, the interferometric images of the droplets are oriented elliptical regions that feature slanted fringes. A sample region is highlighted by the box in Fig.~\ref{Fig:AIPI Data reduction}(a) and is enlarged and shown in Fig.~\ref{Fig:AIPI Data reduction}(b). The Canny edge detection algorithm in MATLAB{\textregistered} was used to obtain the boundary of the elliptical regions, with a sample ellipse shown in Fig.~\ref{Fig:AIPI Data reduction}(b). The centres of the elliptical regions were obtained fitting an ellipse to the detected elliptical regions using the least square method, and then finding the centres of the fitted ellipses. It is important to note that a discrepancy exists between the true $x$ and $y$ coordinates of the droplets and the centre of the ellipses. This discrepancy was obtained and corrected, with the correction procedure presented in Appendix~A.

The depth of the droplets (the droplets $z$ coordinate) was obtained using the formulation proposed in~\cite{wen2021characterization} and is given by  
\begin{equation}
    \label{Eq:zlocation}
    \frac{c-b-z}{c-a-z} = -\frac{M_1}{M_2}\frac{\tan(\alpha)}{\tan(\alpha_0)}.
\end{equation}
In Eq.~(\ref{Eq:zlocation}), $a$ and $b$ are the distances between the surface of the bandpass filter shown in Fig.~\ref{Fig:Diagnostics}(a) and the focal planes 1 and 2, respectively. In Eq.~(\ref{Eq:zlocation}), $c=210~\mathrm{mm}$ is the distance between the origin of the coordinate system and the surface of the bandpass filter. Please note that the choice of bandpass filter surface for the calculations is arbitrary and choosing other reference planes does not influence the values of $z$. This is because $c-a$ and $c-b$ that appear in Eq.~(\ref{Eq:zlocation}) are the distances between the focal planes and the origin of the coordinate system which are fixed in the experiments. In Eq.~(\ref{Eq:zlocation}), $M_1$ and $M_2$ are the magnification ratios for imaging at planes 1 and 2, respectively. Separate experiments (with laser turned off and in the absence of spray and air flow) were performed to obtain the above parameters, with details of these experiments discussed in Appendix~B. In the present study, the values of $a$, $b$, $M_1$, and $M_2$ were 160~mm, 200~mm, 0.53, and 0.54, respectively. $\alpha_0$ is the angle that the power meridian axis of the cylindrical lens (CL1) makes with the $x-z$ plane. This angle was set to $\pi/4$ using the CL1 mount. In Eq.~(\ref{Eq:zlocation}), $\alpha$ depends on the orientation of the fringe patterns, with details of calculation discussed below.

To calculate $\alpha$, first the angle between the slanted fringes, see for example those in Fig.~\ref{Fig:AIPI Data reduction}(b), and the $x$ axis was obtained. This angle is referred to as $\gamma$. To calculate $\gamma$, first, a rectangular region highlighted with the red square in Fig.~\ref{Fig:AIPI Data reduction}(b) was selected. This region is enlarged and shown in Fig.~\ref{Fig:AIPI Data reduction}(c). Then, a two-dimensional Fast Fourier Transform was applied to the data in this region, with the results presented in Fig.~\ref{Fig:AIPI Data reduction}(d). To help elaborating the calculation of $\alpha$, the axes of the power meridian ($x''$) and axis meridian ($y''$) of CL1 were overlaid on Fig.~\ref{Fig:AIPI Data reduction}(d). As can be seen in Fig.~\ref{Fig:AIPI Data reduction}(d), two peaks (see the blue dots) are identified. The angle between the line that connects these two peaks (the brown dashed line) and the $f_y$ axis is referred to as $\gamma$~see~\cite{Gonzales2018}. The angle between this line and the $y''-$axis is referred to $\alpha$. The variation of $\alpha$ versus $z$ was obtained using Eq.~(\ref{Eq:zlocation}) and is presented by the solid red curve in Fig.~\ref{Fig:AIPI Data reduction}(e). As can be seen, the variation of $\alpha$ features an abrupt change at $z=10$~mm. The location of this abrupt change corresponds to focal plane 2, and is due to the change in the sign of $\tan(\alpha)$ in Eq.~(\ref{Eq:zlocation}). That is, at $z = c-b=10~\mathrm{mm}$, $\tan(\alpha)$ becomes zero. In addition to the predictions from Eq.~(\ref{Eq:zlocation}), separate calibration experiments (detailed in Appendix~B) were performed to experimentally obtain the relation between the depth of the droplets and $\alpha$. The results of these calibration experiments are overlaid on Fig.~\ref{Fig:AIPI Data reduction}(e) using the blue solid circular data points. As can be seen, the results obtained from Eq.~(\ref{Eq:zlocation}) and those from the calibration experiments match, suggesting that the formulation in Eq.~(\ref{Eq:zlocation}) can allow for predicting the droplets depth from $\alpha$.    

\begin{figure}
	\centerline{\includegraphics[width=0.5\textwidth]{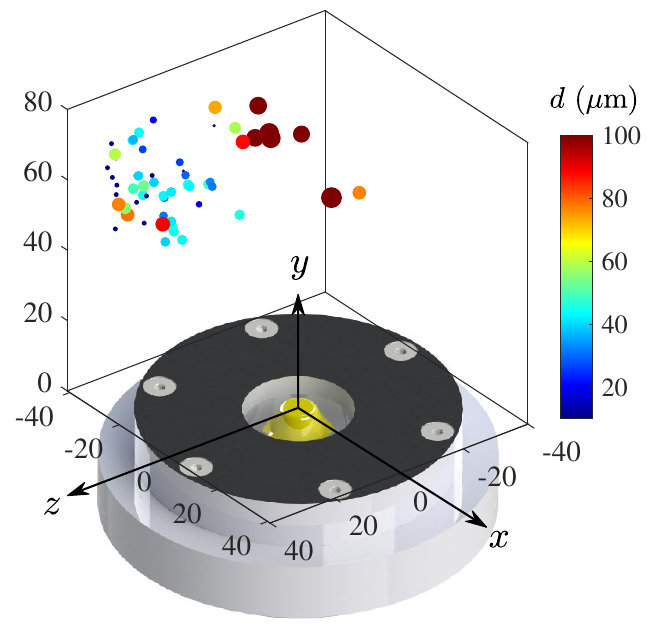}}
	\caption{Sample distribution of the droplets in 3D and the droplets diameter. The results correspond to test condition with $U_\mathrm{b}=6.1~\mathrm{ms^{-1}}$. } 	
	\label{Fig: 3D location and size}
\end{figure}

Once the $z$ location of a droplet was calculated using the above procedure, its diameter was obtained from the formulation proposed in~\cite{wen2021characterization}, which is given by
\begin{equation}
\label{Eq:d}
 d=F(z)\dfrac{2\lambda}{s}\left[\cos\left(\frac{\theta}{2}\right)+\dfrac{m\sin(\frac{\theta}{2})}{\sqrt{m^{2}-2m\cos(\frac{\theta}{2})+1}}\right]^{-1}.
\end{equation}
In Eq.~(\ref{Eq:d}), $F(z)$ represents the dependence of the droplet diameter on its position along the $z$-axis. $F(z)$ is given by
\begin{equation}
\label{Eq:Fz}
F(z)=\dfrac{1}{\sqrt{\left[\dfrac{\cos(\alpha_0)}{M_1(c-a-z)}\right]^2+\left[\dfrac{\sin(\alpha_0)}{M_2(c-b-z)}\right]^2}}.
\end{equation}
In Eq.~(\ref{Eq:d}), $\lambda$ is the wavelength of the AIPI laser, $s$ is the fringe spatial spacing, with a sample shown in Fig.~\ref{Fig:AIPI Data reduction}(c), and $m$ is the index of refraction for distilled water, which is 1.33 as provided in~\cite{hardalupas2010simultaneous}. The uncertainties associated with measuring the droplet depth and diameter were estimated and the details are presented in Appendix~C. These uncertainties depend on the imaging resolution, characteristics of the optics (such as focal length, magnification ratios, and the distance between the focal planes), as well as the characteristics of the fringe patterns (such as their orientation and spacing). It was estimated that the maximum relative uncertainties in estimating the droplet depth and most probable diameter are about 13\% and 27\%, respectively. For all test conditions, the 3D position and the diameter of the droplets were obtained. Figure~\ref{Fig: 3D location and size} shows a representative 3D distribution of the droplets as well as their diameter. The results presented in the figure are for test condition with $U_\mathrm{b}=6.1~\mathrm{ms^{-1}}$. The location of the droplets in 3D was used for identifying the clusters and voids, with relevant details discussed in the following subsection.

\subsection{Clusters and voids identification in the three-dimensional space}
\label{subsection:clusteridentification}
\begin{figure}
	\centerline{\includegraphics[width=0.95\textwidth]{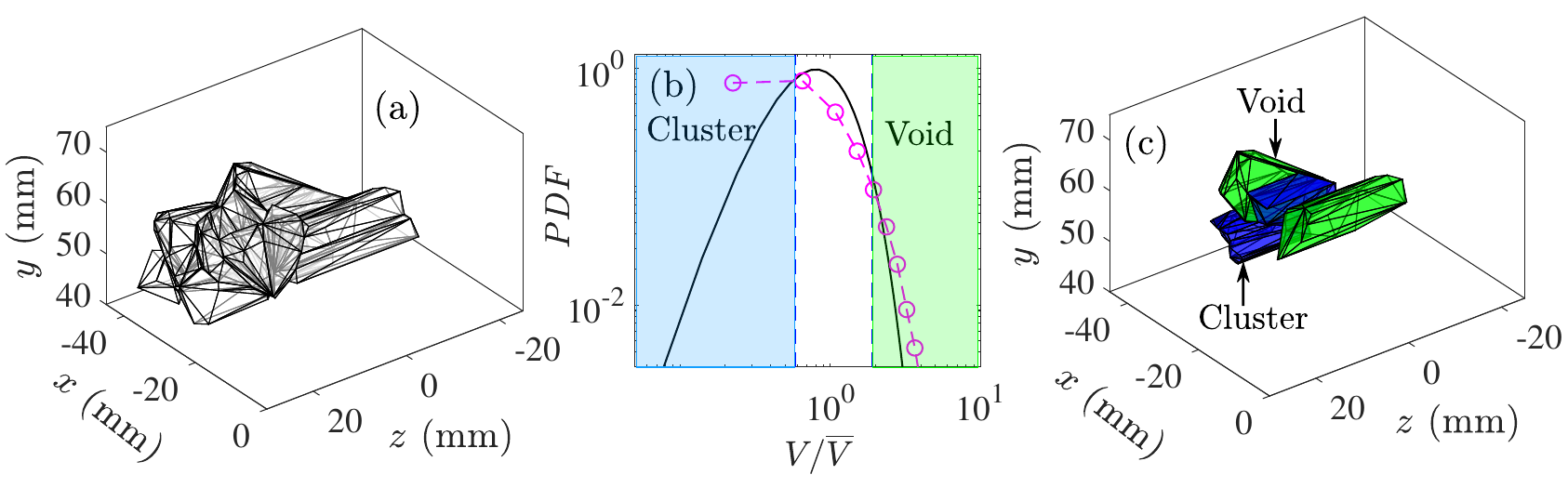}}
	\caption{(a) is the Vorono\"{i} cells generated using the 3D spatial distribution of droplets in Fig.~\ref{Fig: 3D location and size}. (b) presents the PDF of the Vorono\"{i} cells volume normalized by their local mean for test condition with $U_\mathrm{b} =6.1~\mathrm{ms^{-1}}$. Overlaid on (b) is the PDF of a random distribution of the droplets (see the black solid curve). (c) presents the identified clusters (blue) and voids (green).}
	\label{Fig:Data reduction- cluster}
\end{figure}

The spatial distribution of the droplets in 3D space was used to identify the clusters and voids. Using the 3D position of the droplets centre along with the "Voronoin" function in MATLAB{\textregistered}, the 3D Vorono\"{i} cells were obtained first. This was followed by removing the cells intersecting with the boundaries of the AIPI Volume-of-interest. As a representative example, for the droplets identified and presented in Fig.~\ref{Fig: 3D location and size}, the Vorono\"{i} cells were obtained, those intersecting with the AIPI region-of-interest boundaries were removed, and the remaining cells are shown in Fig.~\ref{Fig:Data reduction- cluster}(a). It is important to note that, given the discussion in Section~\ref{subsection:3D}, the centre of all droplets with sample ellipses shown in Fig.~\ref{Fig: 3D location and size} can be obtained; however, the diameter of those inside the focal planes 1 and 2 cannot be calculated. Nonetheless, the 3D position of all detected droplets were used to identify clusters and voids. For joint analysis of the droplet diameter and cluster volume, the droplets with detectable diameter that reside within the clusters/voids were used for analysis. 

Once the Vorono\"{i} cells were obtained, the volumes of the cells were calculated using the ``convhulln" function in MATLAB{\textregistered}. Compared to the particle-laden flow studies in which the particles are initially and relatively uniformly distributed in wind tunnels, in many engineering applications, the particles/droplets feature biased distribution in space, and as a result, several regions in the 3D space feature relatively large concentration of the droplets. In the present study, the pressure swirl atomizer creates an inverted cone of droplets. Such distribution of droplets leads to biased Vorono\"{i} cells volume, and this bias was removed following the procedure described below, which is similar to that used in \cite{rostami2023separate} for 2D Vorono\"{i} cells. To this end, the mean of the local Vorono\"{i} cells volume ($\overline{V}(x,y,z)$) was obtained first. To estimate $\overline{V}(x,y,z)$, the AIPI volume-of-interest was divided into several voxels with length, width, and height of $\Delta x = 4~\mathrm{mm}$, $\Delta y=4~\mathrm{mm}$ and $\Delta z=6~\mathrm{mm}$, respectively. Then, the volume of the Vorono\"{i} cells corresponding to the droplets located within each of these voxels was obtained for all frames and then averaged for each test conditions. To identify the clusters, first, the $PDF(V/\overline{V})$ was obtained, with that for the mean bulk flow velocity of 6.1~$\mathrm{ms^{-1}}$ was shown in Fig.~\ref{Fig:Data reduction- cluster}(b) using the circular date symbol. Provided that the spatial distribution of the droplets followed the Random Poison Process, the PDF of the normalized volume of Vorono\"{i} cells, referred to as $PDF_\mathrm{RPP}$, was estimated using that in~\cite{ferenc2007size} and is given by
\begin{equation}
\label{eq:random}
PDF_\mathrm{RPP}(V/\overline{V})=\frac{B^A}{\Gamma(A)}\left(V/\overline{V}\right)^{A-1}\exp^{(-B V/\overline{V})},
\end{equation}
where $A$ and $B$ are fitting parameters with $A=B=5$. In Eq.~(\ref{eq:random}), $\Gamma$ is the gamma-function with $\Gamma(A) = 24$. The variation of $PDF_\mathrm{RPP}$ versus $V/\overline{V}$ is overlaid on Fig.~\ref{Fig:Data reduction- cluster}(b) by the black solid curve and for comparison purposes. Furthermore, using MATLAB{\textregistered}, 2000 synthetic images were generated. These images contained synthetic particles which were randomly distributed in the AIPI volume-of-interest. The number of synthetic droplets was identical to that shown in Fig.~\ref{Fig:Data reduction- cluster}(a). The PDF of $V/\overline{V}$ for the above synthetic particles was obtained, and it was confirmed that the PDF of the Vorono\"{i} cells normalized volume nearly matches the right-hand-side of Eq.~(\ref{eq:random}). As results in Fig.~\ref{Fig:Data reduction- cluster}(b) show, the $PDF_\mathrm{RPP}$ intersects with the experimentally measured $PDF$ at two normalized volumes, which are $V/\overline{V}(x,y,z)$=~0.58 and 1.90. These threshold values were used to categorize the 3D cells shown in Fig.~\ref{Fig:Data reduction- cluster}(a) either as clusters (for $V \leq 0.58\overline{V}(x,y,z)$) or voids (for $V \geq 1.90\overline{V}(x,y,z)$). That is the Vorono\"{i} cells with volume smaller than 0.58~$\overline{V}(x,y,z)$ were labeled as clusters, and Vorono\"{i} cells with volume larger than 1.90~$\overline{V}$ were labeled as voids. This was followed by merging the cells with intersecting faces. Finally, the identified clusters and voids are shown by the blue and green polyhedrons, respectively, in Fig.~\ref{Fig:Data reduction- cluster}(c).


\section{Results}
The results are grouped into four subsections. In the first subsection, the characteristics of the background turbulent swirling air flow are presented. In the second subsection, the influence of the background flow on the dynamics of the spray as well as the diameter of the droplets are analyzed. This subsection is followed by the analysis of the degree of droplet clustering and the length scales of clusters and voids. Finally, the inter-cluster characteristics are investigated.

\subsection{Characteristic of the background turbulent swirling flow}
\label{subsec:non-reactingflow}
\begin{figure}
	\centerline{\includegraphics[width=0.95\textwidth]{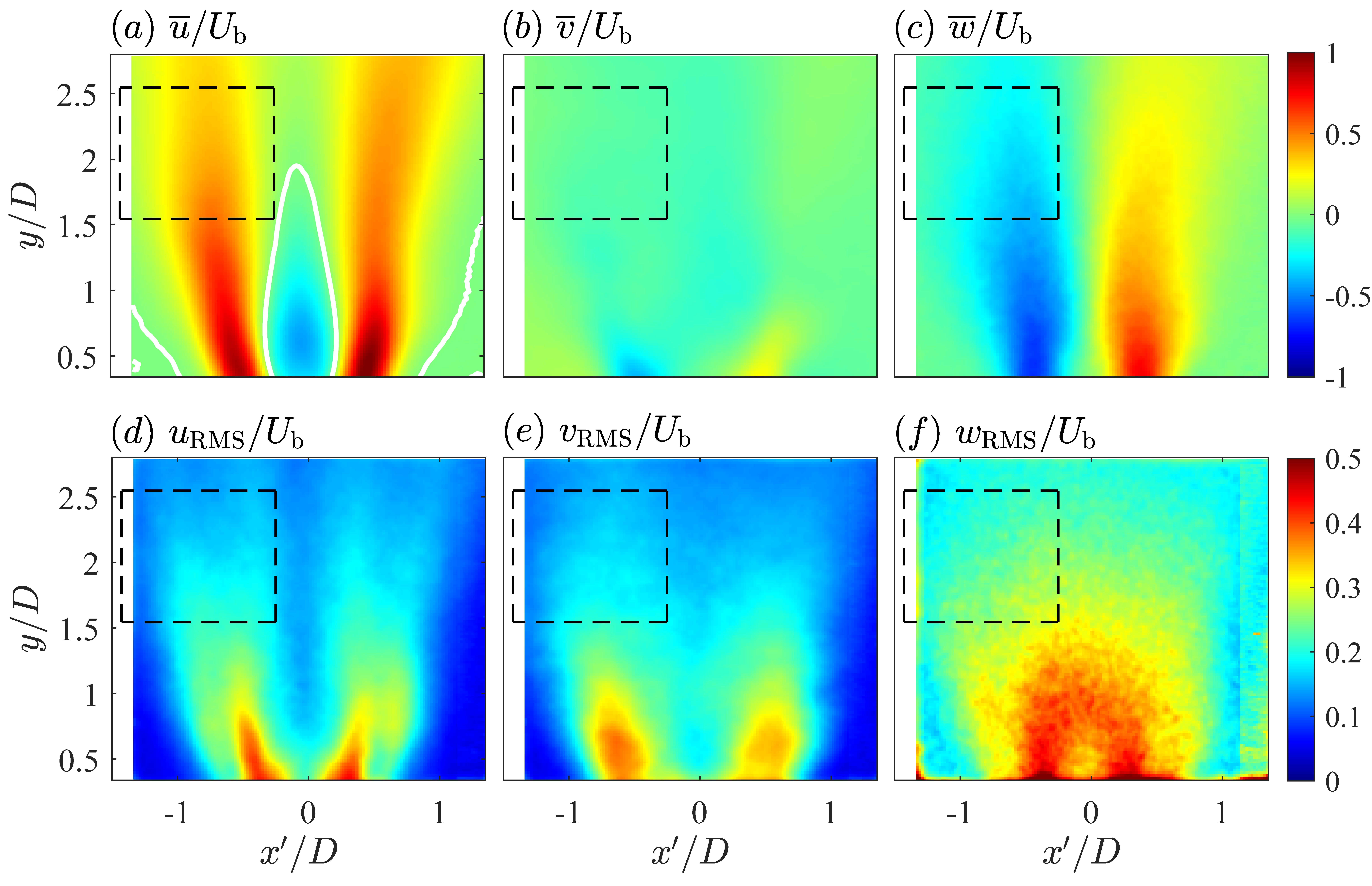}}
	\caption{(a)--(c) are the time-averaged velocity vector components normalized by the mean bulk flow velocity along $y$, $x'$, and $z'$, respectively. Superimposed on (a) are the contours of $\overline{u}/U_\mathrm{b} = 0$ with white colour curves. (d)--(f) present the local RMS of the velocity fluctuations normalized by the mean bulk flow velocity along $y$, $x'$, and $z'$, respectively. The black-dashed box is the intersection of the AIPI region-of-interest and the SPIV plane of measurements. The results pertain to test condition with $U_\mathrm{b}$= 6.1~$\mathrm{ms^{-1}}$.}
	\label{Fig:PIVM}
\end{figure}

Though a summary of the background turbulent air flow characteristics was provided in Table~\ref{tab:tested onditions} to estimate the governing parameters, a detailed analysis is presented here. Specifically, the statistics of the background turbulent velocity fluctuations and the swirl number ($S$) are studied. The local mean velocity and RMS velocity fluctuations were obtained for all test conditions, with the representative results corresponding to $U_\mathrm{b} = 6.1~\mathrm{ms^{-1}}$ shown in Fig.~\ref{Fig:PIVM}. The results in Figs.~\ref{Fig:PIVM}(a--c) present the mean of the velocity vector components normalized by the mean bulk flow velocity along $y$, $x'$, and $z'$, which are $\overline{u}/U_\mathrm{b}$, $\overline{v}/U_\mathrm{b}$, and $\overline{w}/U_\mathrm{b}$, respectively. The results in Figs.~\ref{Fig:PIVM}(d--f) present the local RMS of the velocity fluctuations normalized by the mean bulk flow velocity along $y$, $x'$, and $z'$, which are $u_\mathrm{RMS}/U_\mathrm{b}$, $v_\mathrm{RMS}/U_\mathrm{b}$, and $w_\mathrm{RMS}/U_\mathrm{b}$, respectively. In Fig.~\ref{Fig:PIVM}, the horizontal and vertical axes are normalized by the exit diameter of the nozzle. The intersection of the SPIV FOV and AIPI region-of-interest is overlaid on the figures using the black dashed boxes. In Fig.~\ref{Fig:PIVM}(a), the white curves are the contours of $\overline{u}/U_\mathrm{b} = 0$. As the results in Fig.~\ref{Fig:PIVM}(a) show, there exists a conical vortex breakdown bubble, which is a typical characteristic of swirling flows and has been reported in several past investigations, see for example~\cite{smith2016flow, o2012recirculation, wang2016detailed}. The bubble was present for all test conditions. The results presented in Figs.~\ref{Fig:PIVM}(d--f) show that at a given vertical distance from the nozzle, the RMS of the fluctuating components of the velocity along $x'$ and $y$ features a local minimum at the centre ($x'/D=0$) and local maxima at the shear layers, which are observed and reported in several past investigations, see for example~\cite {wang2016detailed, chen2019large, agostinelli2021impact, kumar2021experimental}. It was confirmed that, for the range of tested mean bulk flow velocities, the mean and RMS of the velocity components normalized by the mean bulk flow velocity feature a similarity behavior, which is similar to the conclusions reported in past studies, see for example \cite{rajamanickam2017insights}.

The swirl number, which is the ratio of the axial flux of the tangential momentum to the product of the axial flux of the axial momentum and $R = D/2$, was estimated for all test conditions. As shown in~\cite{candel2014dynamics}, the above definition of the swirl number leads to the appearance of a pressure dependent term in the formulation of the swirl number, which usually cannot be experimentally estimated. Neglecting this pressure dependent term, the swirl number can be calculated using the formulation in \cite{candel2014dynamics} and is given by
\begin{equation}
\label{Eq: Swirl number}
    S=\dfrac{\\\int_{0}^{R} \overline{w} \ \overline{u}r^{2} \mathrm{d}r}{R\\\int_{0}^{R} \overline{u}^{2}r \mathrm{d}r }.
\end{equation}
For all test conditions, the swirl number was estimated using the velocity data at $y/D = 0.5$ and is about 0.36. Following \cite{degeneve2021impact}, all test conditions feature a relatively moderate swirl number. 

\subsection{Dynamics of the spray formation and droplet characteristics}
\label{subsec:droplet charectristics}
In this subsection, first, the shadowgraphy images are used to study the dynamics of the spray formation. Then, the AIPI data is used to understand the droplet diameter statistics and how these relate to the dynamics of the spray formation. 

\begin{figure}	\centerline{\includegraphics[width=0.95\textwidth]{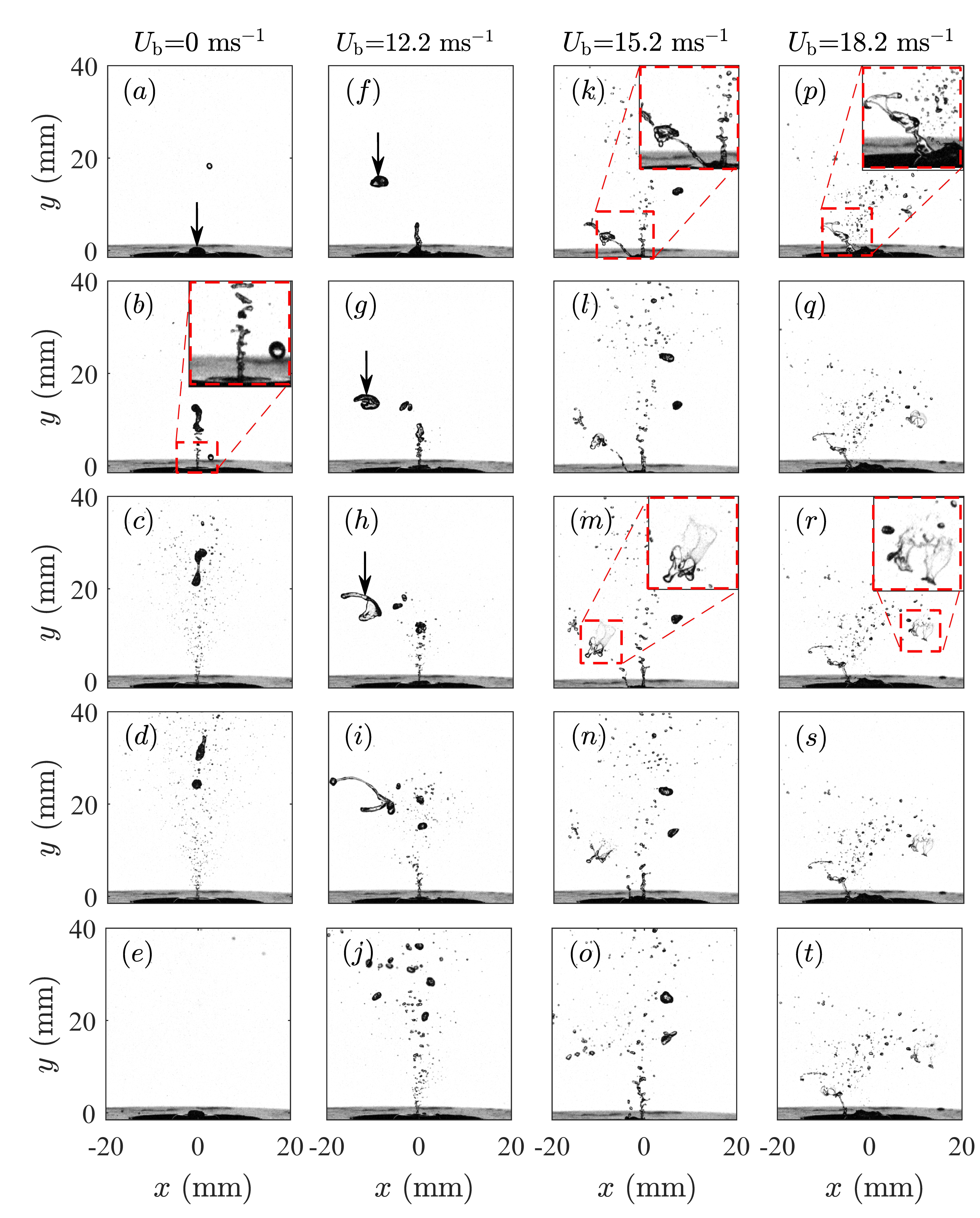}}
\caption{The first to fourth columns present the dynamics of the spray formation for the mean bulk flow velocities of 0, 12.2, 15.2, and 18.2~$\mathrm{ms^{-1}}$, respectively. The videos corresponding to (a--e), (f--j), (k--o), and (p--t) are provided as supplementary materials.}
\label{Fig: Spray formation}
\end{figure}

The shadowgraphy images were analyzed for all test conditions, with the sample results presented in Fig.~\ref{Fig: Spray formation}. The results in Figs.~\ref{Fig: Spray formation}(a--e), (f--j), (k--o), and (p--t) correspond to test conditions with $U_\mathrm{b}$=~0, 12.2, 15.2, and 18.2~$\mathrm{ms^{-1}}$, and present time lapses of $\Delta t = 72.7$, 25.2, 3.9, and 0.9~ms, respectively. The videos corresponding to the results in Fig.~\ref{Fig: Spray formation} are also provided as supplementary materials. As shown in the first column of Fig.~\ref{Fig: Spray formation} and in the corresponding video, for $U_\mathrm{b} = 0~\mathrm{ms^{-1}}$, a bulge of liquid is formed at the tip of the injector initially, see the arrow in Fig.~\ref{Fig: Spray formation}(a). As the time is lapsed, this bulge of liquid stretches and detaches from the injector tip. Concurrent with this, a liquid column emerges from the injector, with the corresponding inset presented in Fig.~\ref{Fig: Spray formation}(b). The bulge of liquid is then stretched and broken into two large droplets; and, the thin liquid column atomizes into small droplets, forming a conical spray. The formation of the large droplets and spray are shown in Figs.~\ref{Fig: Spray formation}(c and d). After this, as shown in Fig.~\ref{Fig: Spray formation}(e), the formation of the spray ceases and a bulge of liquid emerges at the tip of the injector, repeating the process presented in Figs.~\ref{Fig: Spray formation}(a--e).

The presence of the swirling flow at relatively small mean bulk flow velocities ($U_\mathrm{b} \leq 12.2~\mathrm{ms^{-1}}$) does not significantly influence the formation of the conical spray discussed above; however, the liquid bulge undergoes a break-up that is different than that at $U_\mathrm{b}= 0~\mathrm{ms^{-1}}$. Specifically, once the liquid bulge is detached from the injector tip, it is carried into the swirling flow, elongated, and broken into several small droplets. A sample of such liquid bulge that undergoes the above process is highlighted by the arrows in Figs.~\ref{Fig: Spray formation}(f--i). Further increasing the mean bulk flow velocity to $U=15.2~\mathrm{ms^{-1}}$ changes the spray formation. Specifically, it is observed that the liquid bulge (which detached from the tip of the injector for $U_\mathrm{b} \leq 12.2~\mathrm{ms^{-1}}$) remains attached, forming a thin liquid pool at the injector tip and for $U_\mathrm{b} = 15.2~\mathrm{ms^{-1}}$. The interaction of this liquid pool with the swirling flow leads to the formation of elongated structures, with a sample enlarged and shown in the inset of Fig.~\ref{Fig: Spray formation}(k). These structures break up into large droplets which undergo bag break-up as shown in  Fig.~\ref{Fig: Spray formation}(m), similar to that reported in~\cite{ade2023droplet}. For mean bulk flow velocities larger than 15.2~m/s, the formation of the central spray shown in Figs.~\ref{Fig: Spray formation}(a--o) is suppressed, but the liquid pool at the tip of the injector is present and undergoes more frequent formation and shedding of large droplets. These droplets undergo double bag break-up during interaction with the swirling flow (see the inset of Fig.~\ref{Fig: Spray formation}(r)), which is similar to that reported in \cite{ade2023droplet}. 

\begin{figure}
\centerline{\includegraphics[width=1\textwidth]{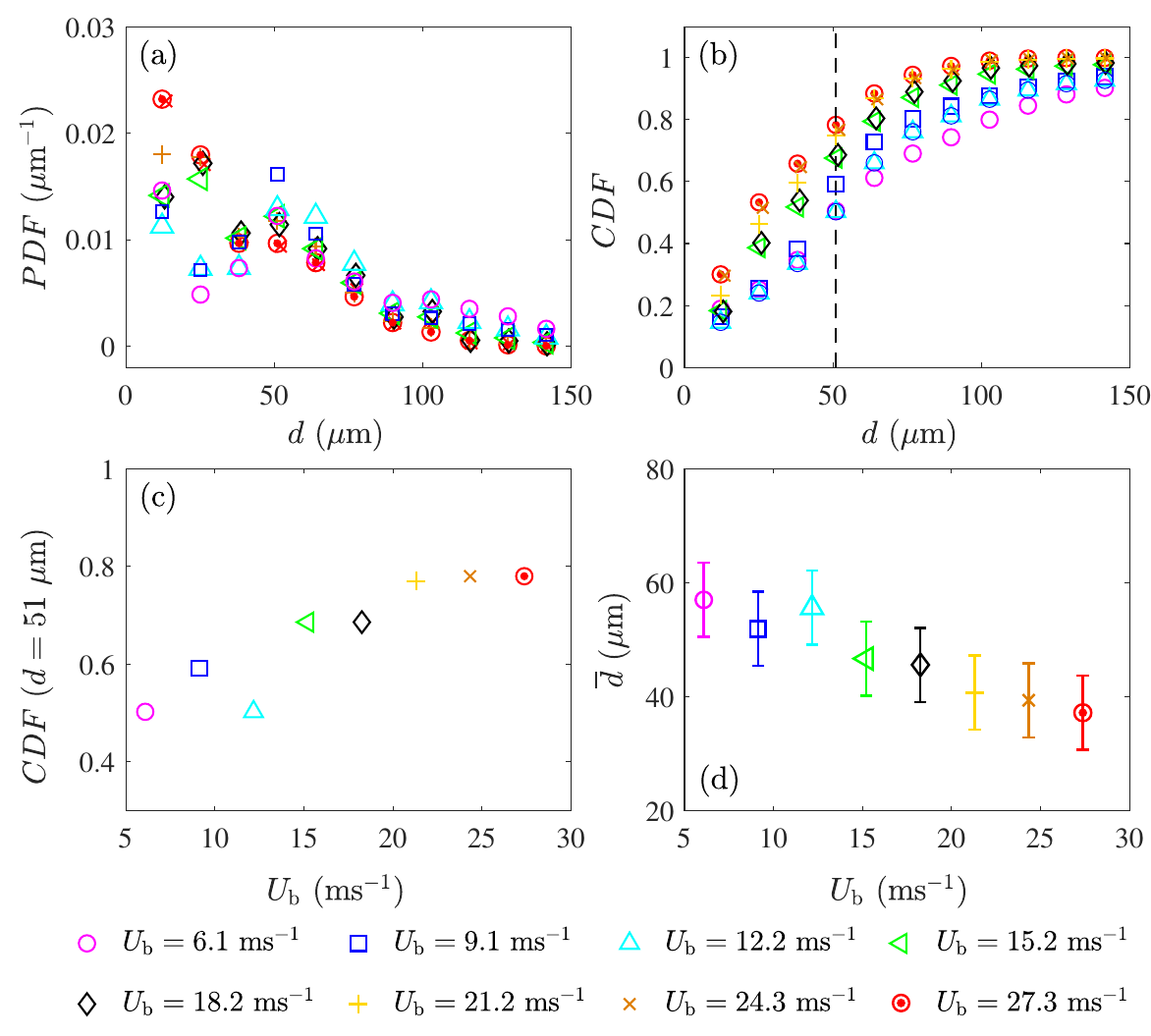}}
\caption{(a) and (b) are the probability and cumulative density functions of the droplet diameter, respectively. (c) presents the variation of the cumulative density function at $d=51~\mu$m versus the mean bulk flow velocity. (d) is the variations of mean droplet diameter versus the mean bulk flow velocity.} 
\label{Fig: Droplet size PDf}
\end{figure}

It is of interest to understand the role of the above described dynamics on the statistics of the droplet diameter. Figure~\ref{Fig: Droplet size PDf}(a) and (b) present the PDF and the Cumulative Density Function (CDF) of the droplet diameter obtained from the AIPI technique. As can be seen, for all test conditions, the PDFs feature a relatively large probability at about 12~$\mu$m. Also, the PDFs feature relatively large values at $d = 51~\mu \mathrm{m}$, with the test conditions corresponding to $U_\mathrm{b} < 24.3~\mathrm{ms^{-1}}$ featuring a local maximum at this droplet diameter. It is observed that increasing the mean bulk flow velocity results in the gradual disappearance of the local maximum present at about $d=51~\mu \mathrm{m}$. To illustrate and quantify this, the values of the CDFs at $d=51~\mu \mathrm{m}$ were presented versus the mean bulk flow velocity in Fig.~\ref{Fig: Droplet size PDf}(c). Increasing $U_\mathrm{b}$ from 6.1 to 12.2 changes $CDF(d = 51~\mu \mathrm{m})$ from about 0.50 to 0.59; however, increasing the mean bulk flow velocity from about 12.2 to 15.2~$\mathrm{ms^{-1}}$ leads to a larger increase of $CDF(d = 51~\mu \mathrm{m})$ from about 0.50 to 0.69. Further increasing the mean bulk flow velocity gradually increases the total number of droplet as shown in Fig.~\ref{Fig: Droplet size PDf}(c). It is hypothesized that the reason for the significant increase in the number of the spray droplets at $U_\mathrm{b} = 15.2~\mathrm{ms^{-1}}$ is due to the (double) bag break-up, with exemplary dynamics presented in Figs.~\ref{Fig: Spray formation}(k--o). For all test conditions, the mean droplet diameter ($\overline{d}$) was obtained and presented in Fig.~\ref{Fig: Droplet size PDf}(d). The length of the error bars in Fig.~\ref{Fig: Droplet size PDf}(d) present the bin size (13~$\mu$m) used for calculating the PDFs and CDFs. As can be seen, the mean diameter of the droplet decreases from about 57 to 37~$\mu$m by increasing the mean bulk flow velocity from 6.1 to 27.3~$\mathrm{ms^{-1}}$.

\subsection{The characteristics of clusters and voids }
In this subsection, the droplets degree of clustering as well as the length scales of clusters and voids are presented. 

\subsubsection{Degree of clustering}
\begin{figure}
\centerline{\includegraphics[width=0.75\textwidth]{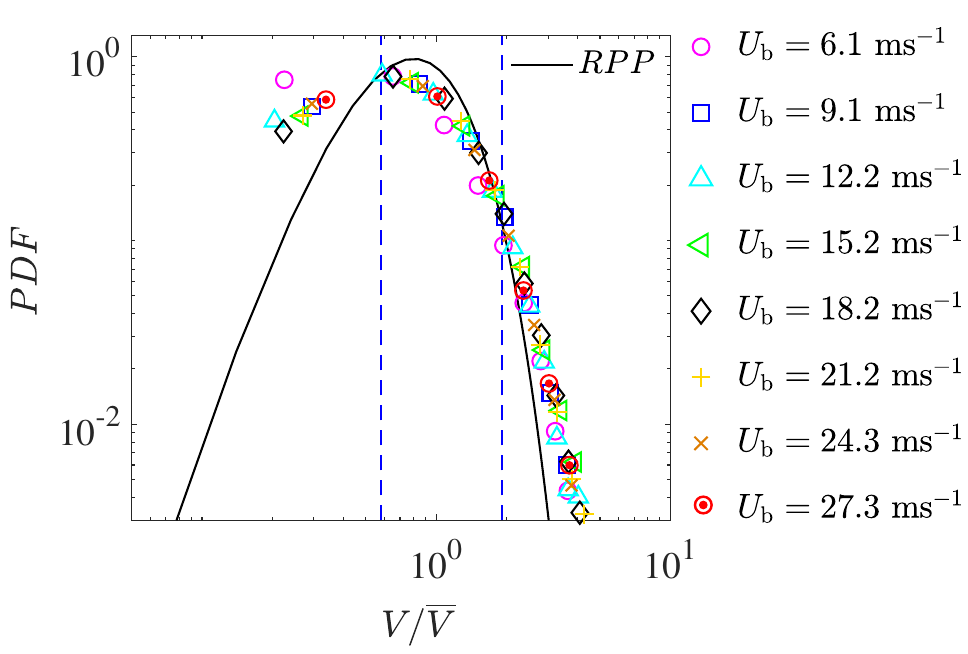}}
\caption{The probability density functions of the Vorono\"{i} cells volume normalized by their local mean value for all test conditions. The PDF of Random Poisson Process is overlaid on the figure with the black solid curve. The blue vertical dashed lines are $V/\overline{V}=0.58$ and $V/\overline{V}=1.90$.}
\label{Fig:PDF of voronoi cells}
\end{figure}

Compared to the studies of \cite{monchaux2010preferential, rostami2023separate}, in which the area of the Vorono\"{i} cells is used for estimating the degree of clustering, in the present study, the volume of the Vorono\"{i} cells is used to calculate the degree of clustering. This approach is similar to that used in the Direct Numerical Simulations of \cite{tagawa2012three}. For all test conditions, the PDFs of the Vorono\"{i} cells volume normalized by their local mean were obtained following the procedure discussed in subsection~\ref{subsection:clusteridentification}, and the results are presented in Fig.~\ref{Fig:PDF of voronoi cells}. As can be seen, the probability density functions of $V/\overline{V}$ collapse for all test conditions. The $PDF_\mathrm{RPP}$ is also overlaid on Fig.~\ref{Fig:PDF of voronoi cells} using the black solid curve. As can be seen, the values of the experimentally obtained PDFs are larger than the $PDF_\mathrm{RPP}$ for $V/\overline{V} < 0.58$ and $V/\overline{V} > 1.90$, and as a result, clustering occurs for all test conditions.

\begin{figure}
\centerline{\includegraphics[width=1\textwidth]{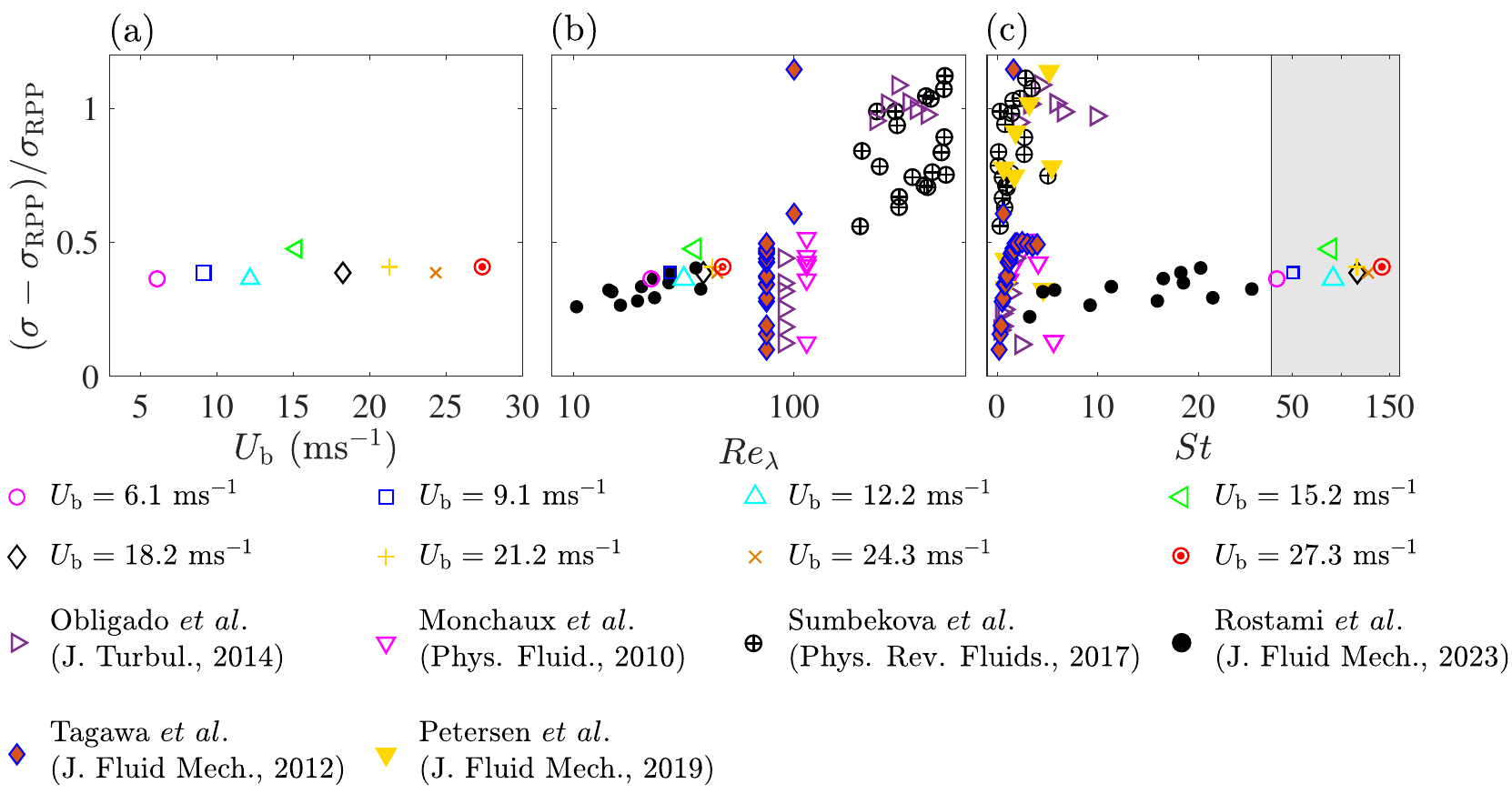}}
\caption{(a), (b), and (c) are the variations of the droplets degree of clustering versus the mean bulk flow velocity, Taylor length scale-based Reynolds number, and Stokes number, respectively. Overlaid on (b) and (c) are the results from \cite{obligado2014preferential,monchaux2010preferential,sumbekova2017preferential,rostami2023separate,tagawa2012three,petersen2019experimental}.}
\label{Fig:Degree of clustering}
\end{figure}

The variation of the degree of clustering versus the mean bulk flow velocity is presented in Fig.~\ref{Fig:Degree of clustering}(a). It is important to note that in calculating the degree of clustering, for particles randomly distributed in 3D, $\sigma_\mathrm{RPP} = 0.45$ (see~\cite{ferenc2007size}), which is different than $\sigma_\mathrm{RPP} = 0.53$ for particles randomly distributed in the 2D pace. Since our analyses are performed for the 3D distribution of particles, $\sigma_\mathrm{RPP}$ was set to 0.45 for calculating the degree of clustering. Our results show that the degree of clustering plateaus at about 0.4. For comparison purposes, the degree of clustering from the present study and those of past investigations are compiled and presented versus $Re_\lambda$ and $St$ in Fig.~\ref{Fig:Degree of clustering}(b) and (c), respectively. To accommodate the presentation of results corresponding to large Stokes numbers, the horizontal axis in Fig.~\ref{Fig:Degree of clustering}(c) was split and different linear scales were used in the figure for results with Stokes numbers smaller and larger than about 34, see the regions with the white and gray backgrounds in the figure. The degree of clustering from 2D measurements of~\cite{obligado2014preferential,monchaux2010preferential,sumbekova2017preferential,petersen2019experimental,rostami2023separate} as well as that obtained from 3D Direct Numerical Simulations of~\cite{tagawa2012three} are overlaid on the region with the white background in Fig.~\ref{Fig:Degree of clustering}(c). The results of the present study are shown on the region with gray background colour in Fig.~\ref{Fig:Degree of clustering}(c). Comparison of the results presented in our previous work, \cite{rostami2023separate}, with those of the present study shows that, for relatively small Taylor length scale-based Reynolds numbers ($Re_\lambda\lesssim 100$) and for a wide range of Stokes numbers ($ 3 \lesssim St \lesssim 142$), increasing $St$ and $Re_\lambda$ slightly increases the degree of clustering from about 0.2 to 0.4. However, the sensitivity of the degree of clustering to Stokes number is larger at larger Taylor length scale-based Reynolds numbers. For example, the results of~\cite{sumbekova2017preferential} show that increasing $Re_\lambda$ from about 200 to 400 increases the degree of clustering from about 0.5 to 1.2. At a fixed Taylor length scale-based Reynolds number ($Re_\lambda \approx 100$), results presented in \cite{monchaux2010preferential, tagawa2012three, obligado2014preferential} suggest that the variation of the degree of droplets clustering versus $St$ features a peak at $St\approx 2-4$. 

\subsubsection{Length scale of clusters and voids}
\begin{figure}
\centerline{\includegraphics[width=0.9\textwidth]{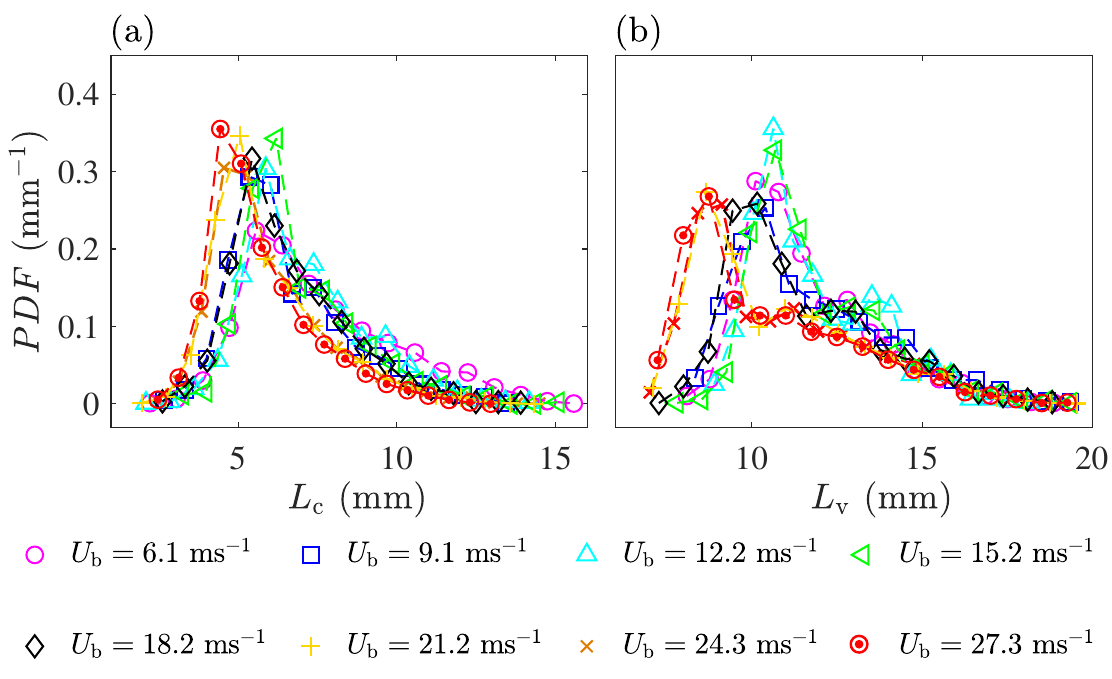}}
\caption{(a) and (b) are the probability density functions of the cluster and void length scales, respectively.}
\label{Fig:Length scale}
\end{figure}

Following the procedure discussed in subsection~\ref{subsection:clusteridentification}, the clusters and voids were identified using the threshold values of $V/\overline{V}$=~0.58 and 1.90 (see the blue dashed lines in Fig.~\ref{Fig:PDF of voronoi cells}). Then, the clusters and voids length scales were estimated using $L_\mathrm{c} = \sqrt[3]{V_\mathrm{c}}$ and $L_\mathrm{v} = \sqrt[3]{V_\mathrm{v}}$, respectively, with $V_\mathrm{c}$ and $V_\mathrm{v}$ being the cluster and void volumes. For all test conditions, the PDFs of $L_\mathrm{c}$ and $L_\mathrm{v}$ are presented in Figs.~\ref{Fig:Length scale}(a) and (b), respectively. As can be seen, the PDFs of the cluster length scale feature a peak, and the PDFs nearly collapse for $L_\mathrm{c} \gtrsim 10$~mm. It is observed that increasing the mean bulk flow velocity from 6.1 to 27.3~$\mathrm{ms^{-1}}$ decreases the most probable value of $L_\mathrm{c}$ from about 5.5 to 4.5~mm. Similar behavior is also observed for the PDF of the voids length scales. Specifically, the PDFs of this parameter nearly collapse for $L_\mathrm{v} \gtrsim 15$~mm; and, the most probable value of $L_\mathrm{v}$ decreases from about 10.1 to 8.7~mm increasing $U_\mathrm{b}$ from 6.1 to 27.3~$\mathrm{ms^{-1}}$.

\begin{figure}
\centerline{\includegraphics[width=0.95\textwidth]{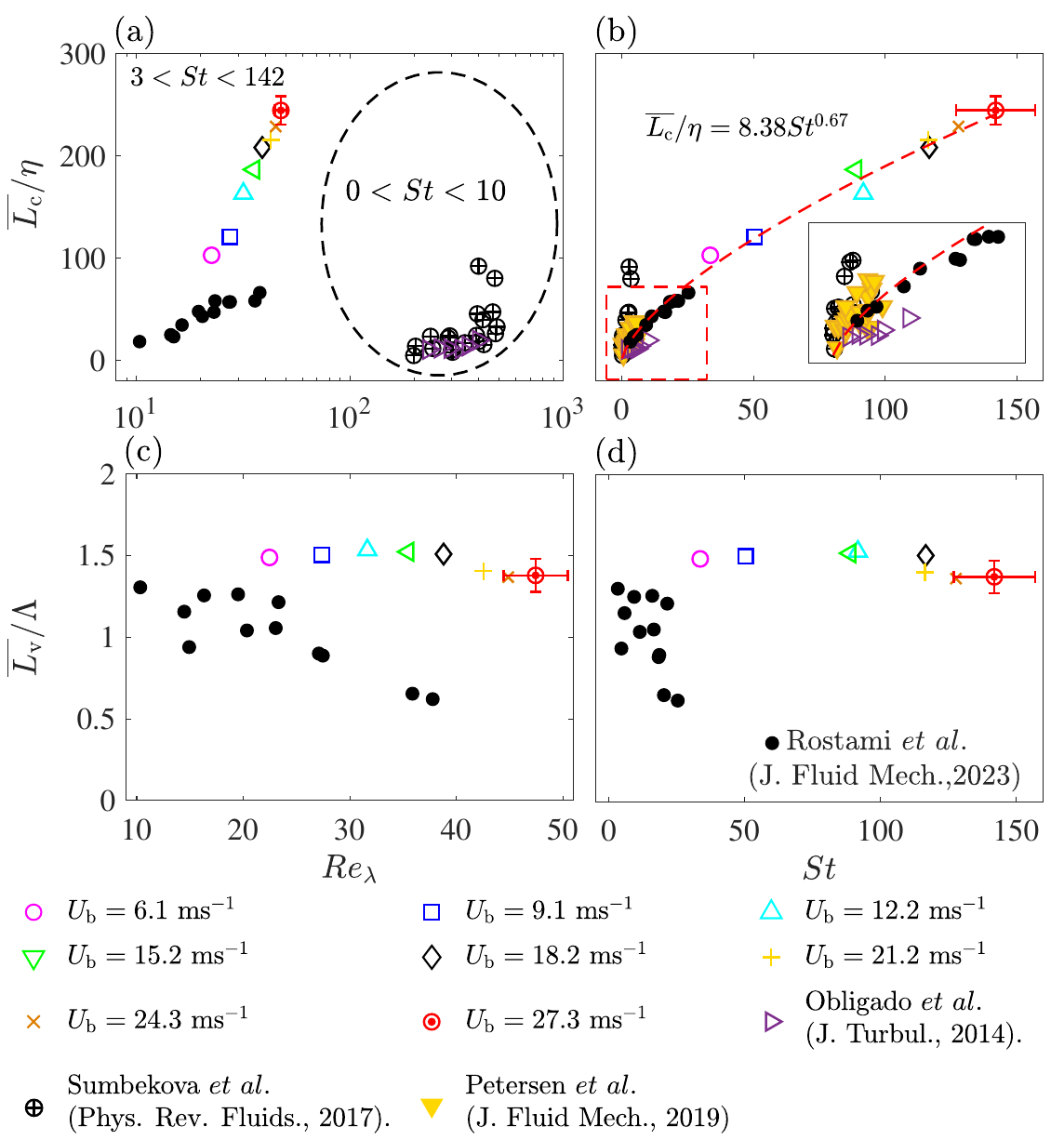}}
\caption{(a) and (b) are the variations of the mean cluster length scale normalized by the Kolmogorov length scale versus $Re_\lambda$ and $St$, respectively. (c) and (d) are the variations of the mean void length scale normalized by the integral length scale versus $Re_\lambda$ and $St$, respectively. Overlaid on (a--d) are the results of \cite{obligado2014preferential}, \cite{sumbekova2017preferential}, \cite{petersen2019experimental}, and \cite{rostami2023separate}.}
\label{Fig:Length scale ratios}
\end{figure}

The majority of past investigations (see for example \cite{obligado2014preferential,sumbekova2017preferential, rostami2023separate}) studied the relations between the mean length scale of the clusters divided by the Kolmogorov length scale as well as how this ratio is influenced by $Re_\lambda$ and/or $St$. To study this, the variations of $\overline{L_\mathrm{c}}/\eta$ versus $Re_\lambda$ and $St$ are presented in Figs.~\ref{Fig:Length scale ratios}(a) and (b), respectively. The lengths of the horizontal error bars in Figs.~\ref{Fig:Length scale ratios}(a and c) as well as (b and d) are the maximum standard deviation in calculating $Re_\lambda$ and $St$ for the black dashed box shown in Fig.~\ref{Fig:PIVM} and from the SPIV data. Also, the lengths of the vertical error bars in Figs.~\ref{Fig:Length scale ratios}(a and c) and (b and d) are the maximum standard deviation in calculating $\overline{L_\mathrm{c}}/\eta$ and $\overline{L_\mathrm{v}}/\Lambda$. For comparison purposes, the results of \cite{obligado2014preferential,sumbekova2017preferential, rostami2023separate} are also overlaid on the figures. The results in Fig.~\ref{Fig:Length scale ratios}(a) show that the relation between $\overline{L}_\mathrm{c}/\eta$ and $Re_\lambda$ depends on the range of the tested Stokes and Taylor length scale-based Reynolds numbers. For relatively large Stokes numbers, the results of~\cite{rostami2023separate} and those of the present study show that increasing the $Re_\lambda$ from about 10 to 47 increases $\overline{L}_\mathrm{c}/\eta$ from about 18 to 244. However, for small Stokes numbers, increasing $Re_\lambda$ from about 200 to 477 increases $\overline{L_\mathrm{c}}/\eta$ from about 5 to 80. Compared to the above and independent of the tested $Re_\lambda$, the results of the present study and those of \cite{rostami2023separate,obligado2014preferential,sumbekova2017preferential,tagawa2012three,petersen2019experimental} shown in Fig.~\ref{Fig:Length scale ratios}(b) suggest that the variation of $\overline{L_\mathrm{c}}/\eta$ versus the Stokes number follows an increasing trend. It is also anticipated that as the Stokes number approaches zero, the cluster size should approach zero. As such, the least-square method was used to fit a power law-relation (with $\eta^{-1}\overline{L_\mathrm{c}}(St = 0) = 0$) to the results in Fig.~\ref{Fig:Length scale ratios}(a). This relation is shown by the red dashed curve and is given by
\begin{equation}
\label{Eq:powerlaw}
    \frac{\overline{L}_\mathrm{c}}{\eta} \approx 8.38St^{0.67}.
\end{equation}
It is important to highlight that the positive relation between the mean cluster length scale normalized by the Kolmogorov length scale and the Stokes number was proposed for $St<20$ in our previous work. In there, it was also suggested that $\overline{L_\mathrm{c}}/\eta$ appear to follow a linear relation with $St$. For the relatively large Stokes numbers examined in the present study, the positive relation in \cite{rostami2023separate} is confirmed. However, a power-law relation in the form of Eq.~(\ref{Eq:powerlaw}) appears to best present the large range of Stokes numbers examined here and in past investigations. 
 
The variations of mean void length scale normalized by the integral length scale ($\overline{L_\mathrm{v}}/\Lambda$) versus both $Re_\lambda$ and $St$ were obtained, and the results are presented in Fig.~\ref{Fig:Length scale ratios}(c) and (d), respectively. The length of the vertical error bar in Figs.~\ref{Fig:Length scale ratios}(c) and (d) is the maximum standard deviation in estimating $\overline{L_\mathrm{v}}/\Lambda$. Overlaid on these figures by the solid black circular data symbol are the values of $\overline{L_\mathrm{v}}/\Lambda$ from the authors' past work. As can be seen, the mean length scales of the voids are on the order of the integral length scale ($\overline{L_\mathrm{v}}\approx 1.5 \Lambda$) for the range of examined $Re_\lambda$ and $St$.

\subsection{Inter-cluster and inter-void characteristics}
The number density of the droplets inside the clusters and voids as well as the joint droplets and clusters characteristics are studied in this subsection.

\subsubsection{Number density of droplets within the clusters and voids}

\begin{figure}
\centerline{\includegraphics[width=1\textwidth]{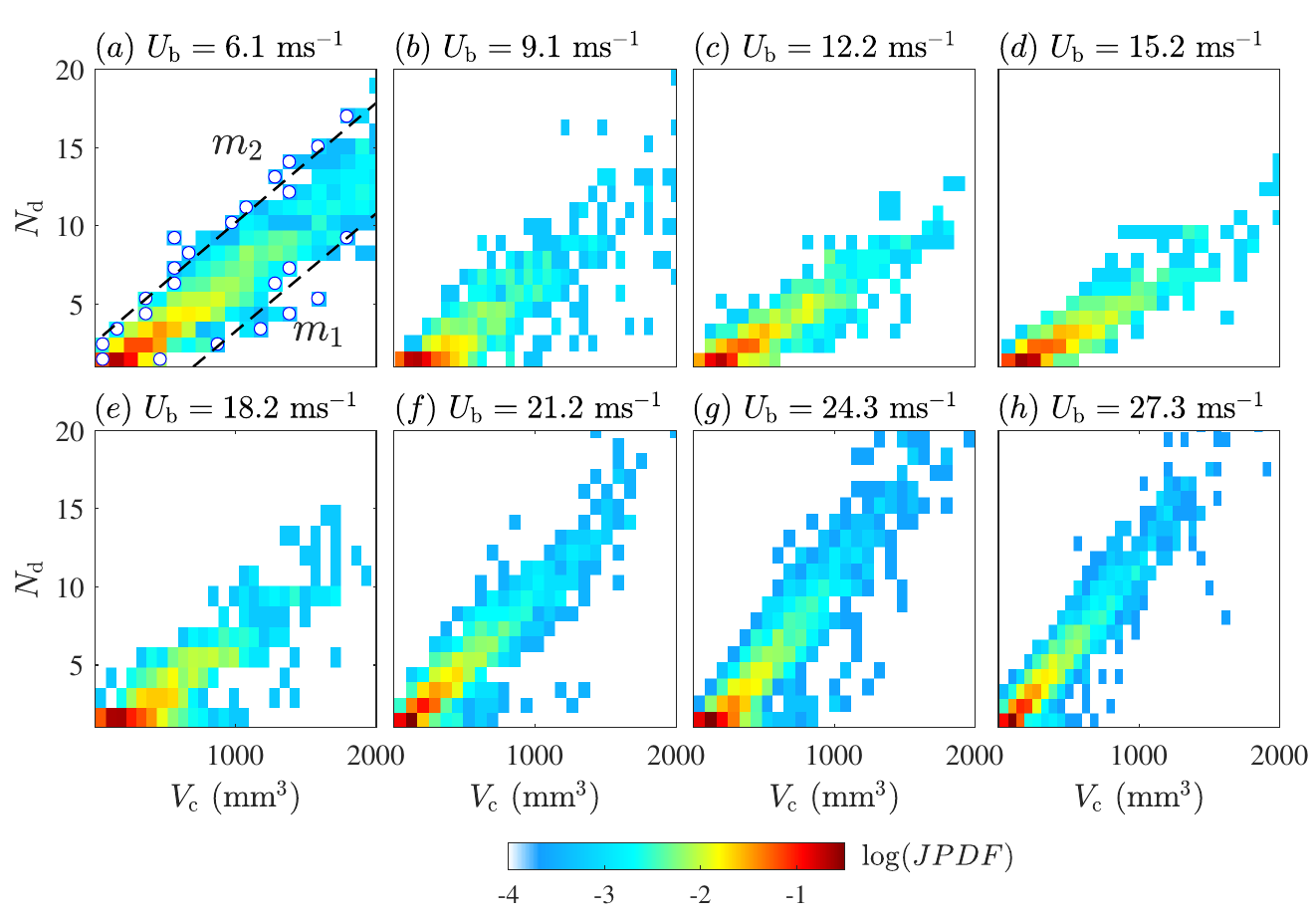}}
\caption{The logarithmic JPDFs of the number of droplets within clusters and the cluster volume. (a)--(h) corresponds to test conditions with $U_\mathrm{b}$= 6.1, 9.1, 12.2, 15.2, 18.2, 21.2, 24.3, and 27.3~$\mathrm{ms^{-1}}$, respectively.}
\label{Fig:Cluster number density}
\end{figure}

\begin{figure}
\centerline{\includegraphics[width=1\textwidth]{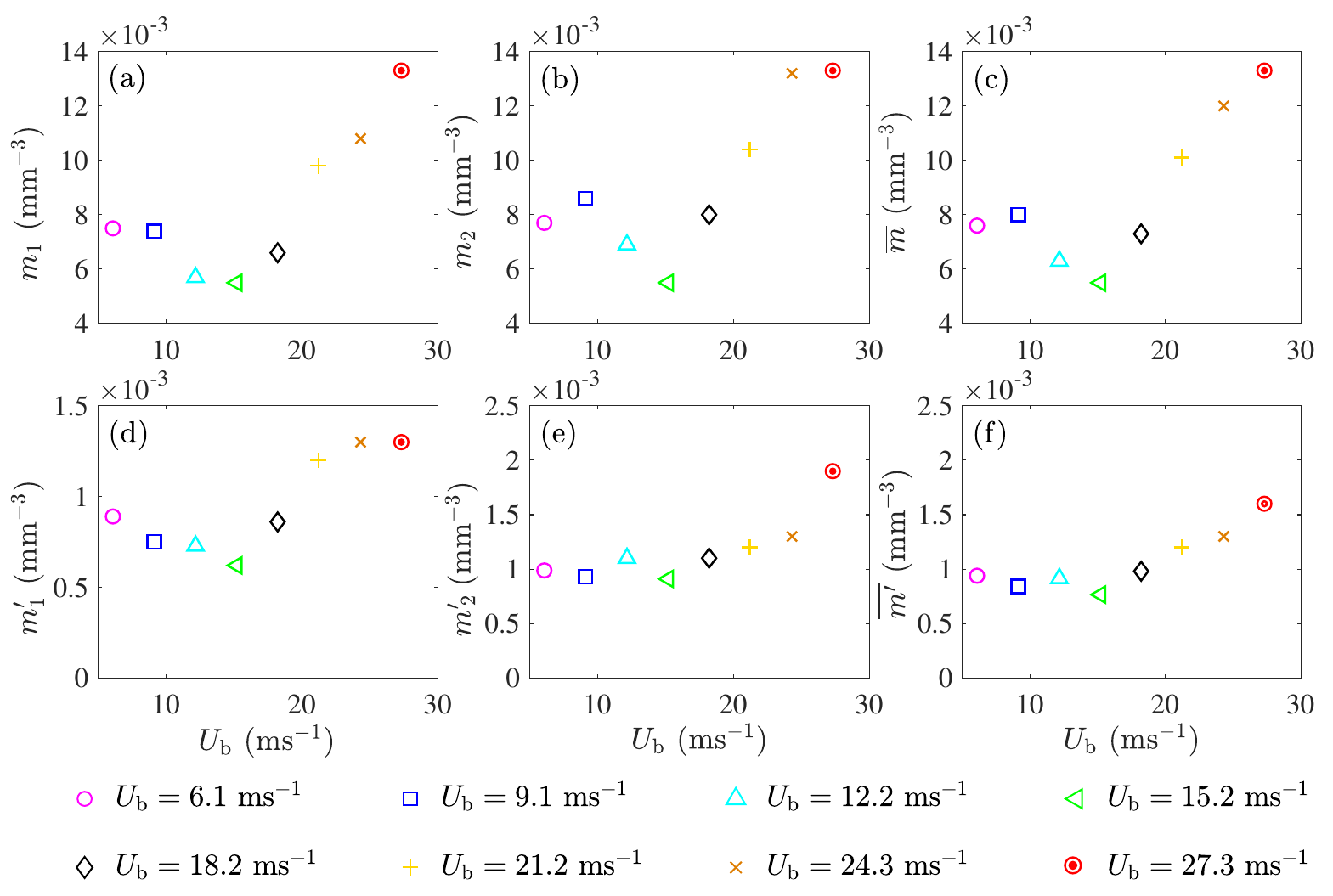}}
\caption{(a)--(c) present the variations of $m_1$, $m_2$, and $0.5(m_1+m_2)$ versus $U_\mathrm{b}$, respectively. (d--f) are the variations of $m^\prime_1$, $m^\prime_2$, and $0.5(m^\prime_1+m^\prime_2)$ versus $U_\mathrm{b}$, respectively.}
\label{Fig:Number density vs U}
\end{figure}

\begin{figure}
\centerline{\includegraphics[width=1\textwidth]{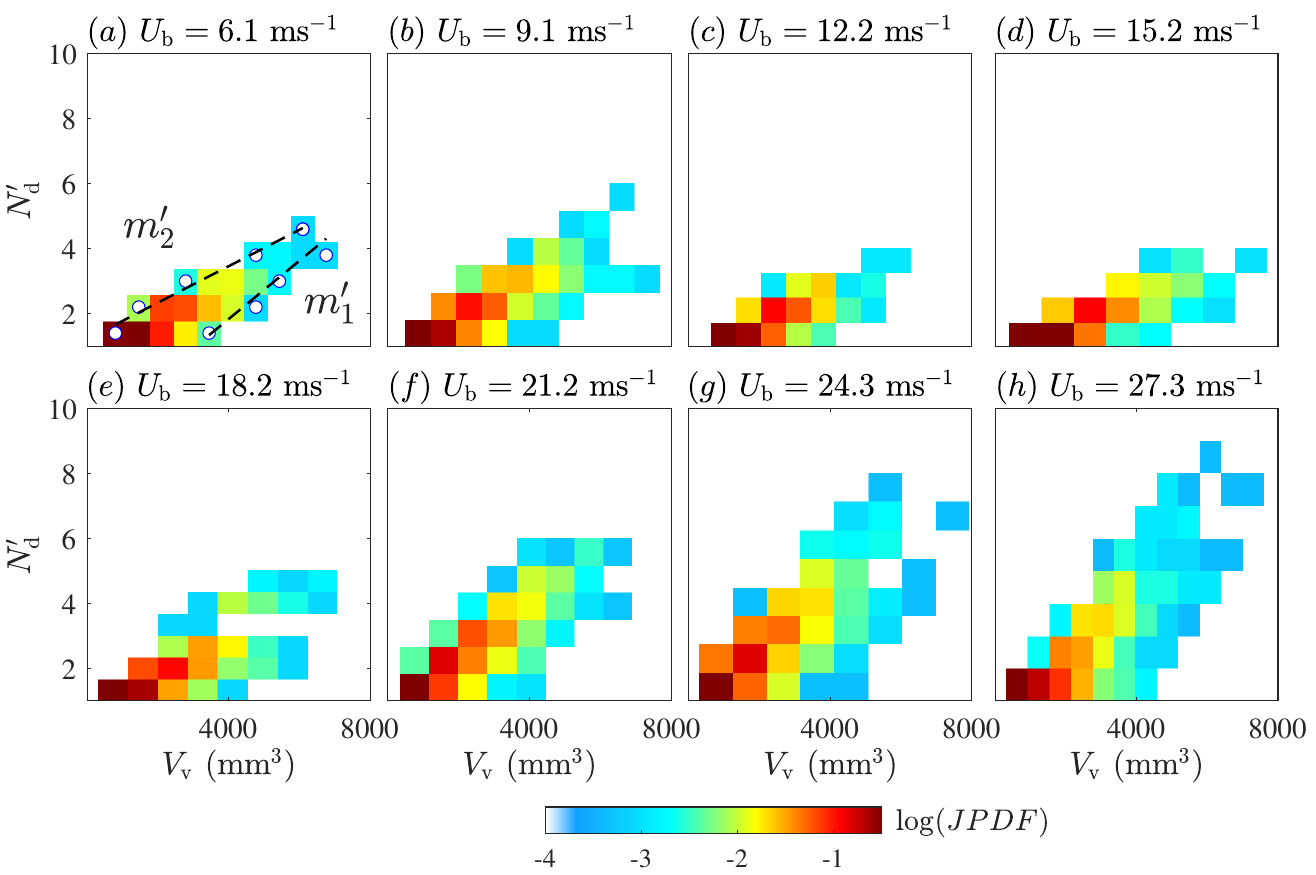}}
\caption{The logarithmic JPDFs of the number of droplets within voids and the void volume. (a)--(h) are for test conditions with $U_\mathrm{b}$= 6.1, 9.1, 12.2, 15.2, 18.2, 21.2, 24.3, and 27.3~$\mathrm{ms^{-1}}$, respectively.}
\label{Fig:Void number density}
\end{figure}

Figure~\ref{Fig:Cluster number density} presents the Joint Probability Density Function (JPDF) of the number of droplets ($N_\mathrm{d}$) that reside within the clusters and the volume of the clusters ($V_\mathrm{c}$). The JPDFs are presented in a logarithmic scale to improve the clarity of presentations. The results in Figs.~\ref{Fig:Cluster number density}(a--h) correspond to test conditions with the man bulk flow velocities of 6.1, 9.1, 12.2, 15.2, 18.2, 21.2, 24.3, and 27.3~$\mathrm{ms^{-1}}$, respectively. As can be seen, there exists a positive relation between the number of droplets and the volume of the clusters, i.e. larger clusters accommodate more number of droplets. It can also be seen that increasing the mean bulk flow velocity influences the above relation. To quantify this, similar to \cite{rostami2023separate}, the combination of ($N_\mathrm{d}, V_\mathrm{c}$) data points at which the JPDF remarkably changes by varying $V_\mathrm{c}$ at a fixed $N_\mathrm{d}$ were obtained, with representative results shown by the white circular data points in Fig.~\ref{Fig:Cluster number density}(a). Then, two lines with slopes of $m_1$ and $m_2$ were fit to the white data points using the least square technique. Then, the average number density was estimated using $\overline{m} = 0.5(m_1+m_2)$. The variations of $m_1$, $m_2$, and $\overline{m}$ versus the mean bulk flow velocity are presented in Figs.~\ref{Fig:Number density vs U}(a)--(c), respectively. As can be seen, increasing the mean bulk flow velocity from 6.1 to 15.2~$\mathrm{ms^{-1}}$ decreases $m_1$, $m_2$, and $\overline{m}$. The decrease in the number density of the clusters is similar to the observation reported in \cite{rostami2023separate} for sprays injected in turbulent co-flows. However, for test conditions with  $U_\mathrm{b}\gtrsim 15.2~\mathrm{ms^{-1}}$, $m_1$, $m_2$, and $\overline{m}$ increase with increasing the mean bulk flow velocity. Similar to the above analysis, for all test conditions, the JPDF of the number of droplets inside a given void ($N_\mathrm{d}^{\prime}$) with volume $V_\mathrm{v}$ were obtained, and the corresponding results are presented in Fig.~\ref{Fig:Void number density}. As can be seen, similar to the results presented in Fig.~\ref{Fig:Cluster number density}, a positive relation exists between $N_\mathrm{d}^{\prime}$ and $V_\mathrm{v}$, with the slope of this relation depending on the mean bulk flow velocity. Similar to the procedure discussed above for calculation of the droplets number density inside the clusters, those for the voids ($m^\prime_1$, $m^\prime_2$, and $\overline{m^\prime}= 0.5(m^\prime_1+m^\prime_2)$) were calculated, and the sample results are shown in Fig.~\ref{Fig:Void number density}(a). For all test conditions, the above parameters were obtained and their variation versus the mean bulk flow velocity are presented in Fig.~\ref{Fig:Number density vs U}(d)--(f). As can be seen, similar to the variations of the number density of the droplets that reside inside the clusters (see Figs.~\ref{Fig:Number density  vs U}(a)--(c)), $m\prime_1$, $m\prime_2$, and $\overline{m\prime}$ attain minimum values at $U_\mathrm{b}=15.2~\mathrm{ms^{-1}}$. Comparison of the results presented in Figs.~\ref{Fig:Number density  vs U}(a)--(c) with those in Figs.~\ref{Fig:Number density  vs U}(d)--(f) shows that the number density of droplets that reside within the clusters is about one order of magnitude larger than the number density of droplets that reside within the voids. This is similar to the conclusions presented in~\cite{rostami2023separate}; however, the results of the present study extends those in \cite{rostami2023separate} to large Stokes numbers that are of relevance to engineering applications such as gas turbine engine combustors.


\begin{figure}
\centerline{\includegraphics[width=0.65\textwidth]{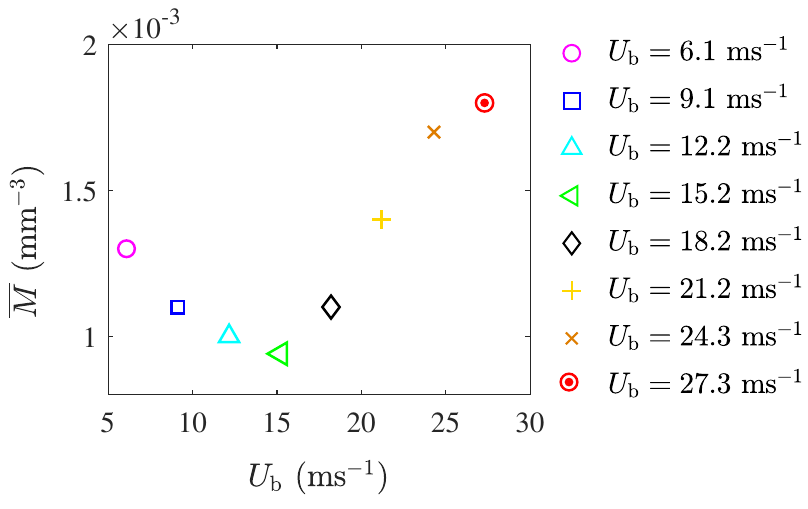}}
\caption{The mean number density of the droplets versus the mean bulk flow velocity.}
\label{Fig:Avergae number density}
\end{figure} 

\begin{figure} 
\centerline{\includegraphics[width=0.9\textwidth]{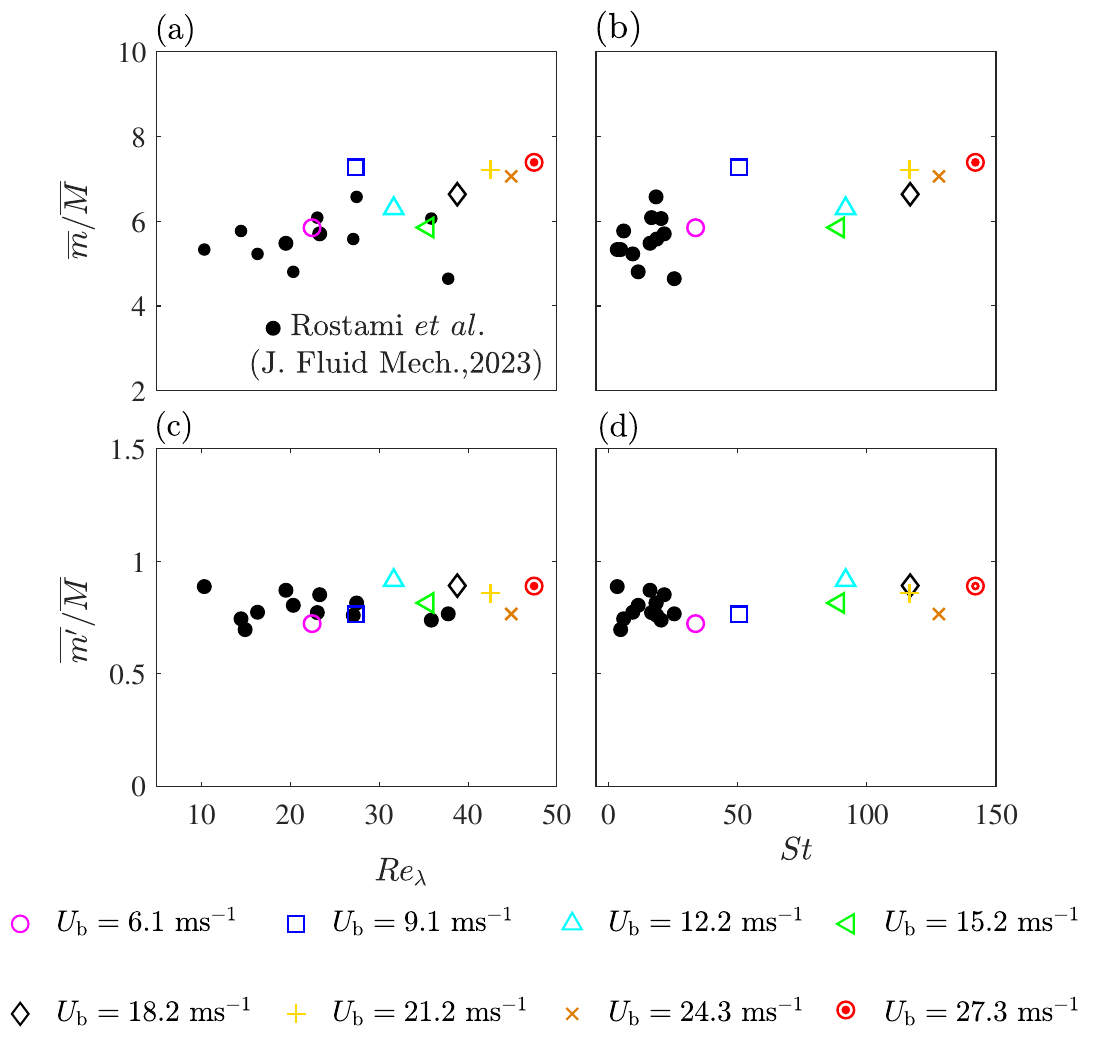}}
\caption{(a) and (b) are the variations of the mean number density of droplets within clusters divided by the mean global number density versus $Re_\lambda$ and $St$, respectively. (c) and (d) are the variations of the mean number density of droplets within voids divided by the mean global number density versus $Re_\lambda$ and $St$, respectively. Overlaid in (a--d) are the results of \cite{rostami2023separate}.}
\label{Fig:Number density ratios}
\end{figure}

It is of interest to investigate the reason for the number density of the droplets within both clusters and voids featuring a minimum at the mean bulk flow velocity of 15.2~$\mathrm{ms^{-1}}$. To this end, the total number density ($\overline{M}$, the ratio of the total number of droplets identified in the AIPI region-of-interest divided by the volume of this region) was obtained for all test conditions and the results are shown in Fig.~\ref{Fig:Avergae number density}. As can be seen, the values of the total number density of the droplets also feature a minimum at $U_\mathrm{b}= 15.2~\mathrm{ms^{-1}}$. For $0 \lesssim U_\mathrm{b} \lesssim 15.2~\mathrm{ms^{-1}}$, increasing the mean bulk flow velocity disperses the particles out of the AIPI region-of-interest, decreasing the total number of the detected droplets. However, for $U_\mathrm{b} \gtrsim 15.2~\mathrm{ms^{-1}}$, increasing the mean bulk flow velocity leads to the occurrence of the (double) bag break-up (see Fig.~\ref{Fig: Spray formation} and the videos in the supplementary materials), which increases the number of generated droplets. It is concluded that the dynamics that influence the droplets generation and atomization influences the trend associated with the variation of $\overline{M}$ versus $U_\mathrm{b}$ which affects the the variations of $\overline{m}$ and $\overline{m^\prime}$ with $U_\mathrm{b}$. 

\cite{rostami2023separate} showed that, though the number density of the droplets that reside within the clusters and voids can vary by changing the background flow characteristics, $\overline{m}/\overline{M}$ and $\overline{m^{\prime}}/\overline{M}$ remain nearly unchanged and about 5.5 and 0.8, respectively. It is of interest to study if the findings of \cite{rostami2023separate} can be extended to sprays injected to turbulent swirling co-flows with relatively large Stokes numbers. As such, the variations of $\overline{m}/\overline{M}$ versus the Taylor length scale-based Reynolds and Stokes numbers are presented in Fig.~\ref{Fig:Number density ratios}(a) and (b), respectively. Also, the variations of $\overline{m^{\prime}}/\overline{M}$ versus $Re_\lambda$ and $St$ are presented in Fig.~\ref{Fig:Number density ratios}(c) and (d), respectively. Additionally, the results of \cite{rostami2023separate} are overlaid on the figures. It can be seen that, for similar values $Re_\lambda$ and $St$, $\overline{m}/\overline{M}$ and $\overline{m^{\prime}}/\overline{M}$ obtained from the present study matches those of our past work. This may suggest that the normalized number densities of the droplets within clusters and voids are independent of the presence of the swirling flow and the type of the utilized diagnostics (2D versus 3D) for $20 \lesssim Re_\lambda \lesssim 40$. The results presented in Fig.~\ref{Fig:Number density ratios}(c) and (d) show that increasing $Re_\lambda$ and $St$ to about 50 and 140 do not change the normalized number density of droplets within voids and this number remains about 0.8. However, the results of the present study and those of \cite{rostami2023separate} show that increasing $Re_\lambda$ from about 10 to 47 and $St$ from about 3 to 142 increases the normalized number density of droplets inside the clusters from about 4.6 to 7.4.

\begin{figure}
\centerline{\includegraphics[width=1\textwidth]{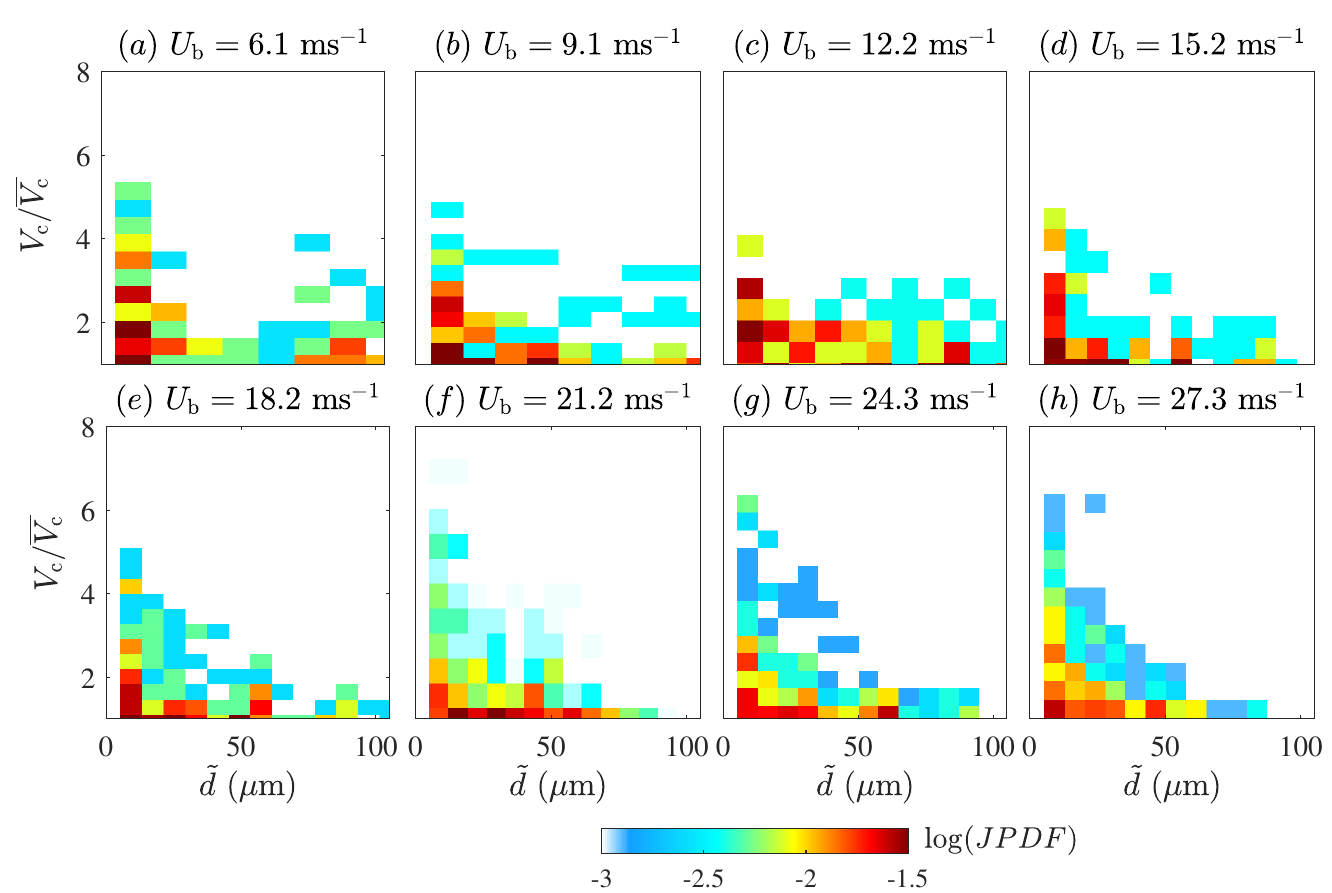}}
\caption{The logarithmic JPDFs of normalized cluster volume and the mean diameter of the droplets residing within the clusters. (a)--(h) correspond to test conditions with $U_\mathrm{b}$= 6.1, 9.1, 12.2, 15.2, 18.2, 21.2, 24.3, and 27.3~$\mathrm{ms^{-1}}$, respectively.}
\label{Fig:cluster JPDF}
\end{figure}

\subsubsection{Joint probability density function of the droplet diameter and cluster volume}
The joint probability density functions of the clusters volume normalized by their mean volume ($V_\mathrm{c}/\overline {V_\mathrm{c}}$) and mean diameter ($\tilde{d}$) of the droplets that reside within a given cluster were obtained for all test conditions, and the results are presented in Fig.~\ref{Fig:cluster JPDF}. The results shown in Figs.~\ref{Fig:cluster JPDF}(a)--(h) pertain to test conditions with the man bulk flow velocities of 6.1, 9.1, 12.2, 15.2, 18.2, 21.2, 24.3, and 27.3~$\mathrm{ms^{-1}}$, respectively. The JPDFs are presented in logarithmic scale to improve the presentation of the results. A similar analysis was also performed in \cite{rostami2023separate} using 2D measurements. Compared to these measurements, for similar total number of droplets, the 3D measurements yield smaller number of droplets in the clusters. As such, the joint PDFs of $V_\mathrm{c}/\overline {V_\mathrm{c}}$ and $\tilde{d}$ are scattered, especially at small mean bulk flow velocities. However, conclusions can be made for mean bulk flow velocities larger than 15.2~$\mathrm{ms^{-1}}$. For these large mean bulk flow velocities, the results show that clusters with large volumes are occupied by droplets with relatively small mean diameter and small variability of their diameter. However, the clusters with relatively small volumes carry droplets with a broader range of diameter. This observation is consistent with the results obtained from 2D measurements reported in~\cite{rostami2023separate}. However, those reported here extend the findings of \cite{rostami2023separate} to larger Stokes numbers that are relevant to engineering applications and using 3D measurements.

\section{Concluding remarks}
The joint droplets and clusters characteristics of large Stokes-number sprays interacting with turbulent swirling air flows were investigated experimentally. A flow apparatus, which facilitated the injection of water spray droplets in a swirling co-flow of air, was employed. The Astigmatic Interferometric Particle Imaging (AIPI) was used for measuring the three-dimensional position as well as the diameter of the spray droplets. Separate shadowgraphy measurements were performed to visualize the influence of background flow on the formation of spray droplets. The characteristics of the background turbulent flow were studied using a separate Stereoscopic Particle Image Velocimetry technique. For all test conditions, the water injection flow rate and the swirl number were fixed at about 14 gr/min and 0.4, respectively. All tested sprays were dilute, with volume fractions ranging from about $0.4\times10^{-6}$ to $0.7\times10^{-6}$. The mean diameter of the droplets varied from about 37~$\mu$m to 57~$\mu$m, changing the Stokes number from about 34 to 142, which is relatively large compared to that of past investigations. The Taylor length scale-based Reynolds number varied from about 22 to 47, rendering our test conditions moderately turbulent compared to that of past studies.   

The shadowgraphy images showed that the presence of the swirling flow significantly changed the spray formation dynamics as well as the break-up of relatively large droplets. Specifically, for mean bulk flow velocities larger than 15.2~$\mathrm{ms^{-1}}$, the (double) bag break-up of large droplets led to their atomization. Analysis of the droplets diameter measured from the AIPI technique showed that, for mean bulk flow velocities smaller than 15.2~$\mathrm{ms^{-1}}$, the Probability Density Function (PDF) of the droplet diameter featured a peak at about 51~$\mu$m. However, once the mean bulk flow velocity increased to values larger than 15.2~$\mathrm{ms^{-1}}$, the above peak of the droplet diameter PDF gradually disappeared. The cumulative distribution function of the droplet diameter and shadowgraphy images showed that the occurrence of the (double) bag break-up leads to the generation of a large number of droplets with diameter smaller than 51~$\mu$m, which smeared out the presence of the local maximum at this diameter and at large mean bulk flow velocities.

The spatial location of the droplets was used to obtain the 3D Vorono\"{i} cells. The PDF of the cells volume normalized by the locally estimated value was used to obtain the degree of clustering as well as the clusters and voids length scales. The diameter of the droplets as well as the identified clusters in 3D were then used to obtained the joint characteristics. The results showed that, for relatively moderate values of the Taylor length scale-based Reynolds number, increasing the Stokes number plateaus the degree of clustering at about 0.4. The PDF of the cluster and void length scales showed that both PDFs featured a peak that is skewed towards small length scales. Specifically, it was obtained that increasing the mean bulk flow velocity from about 6.1 to 27.3~$\mathrm{ms^{-1}}$ decreased the most probable cluster and void length scales from about 5.5 to 4.5~mm and 10.1 to 8.7~mm, respectively. It was concluded that the variation of the mean cluster length scale normalized by the Kolmogorov length scale versus the Stokes number followed a power-law relation, with the exponent of the power law being 0.67. Compared to the normalized mean cluster length scale, the mean void length scale normalized by the integral length scale was nearly insensitive to the test condition and was about 1.5.

The results showed that, although the mean number density of the generated droplets featured a minimum at the mean bulk flow velocity (15.2~$\mathrm{ms^{-1}}$) which was specific to the swirling flows used here, the ratio of the droplets number density residing within the clusters and voids followed trends comparable to our past study in which a turbulent jet co-flow was used. It was concluded that, for moderately turbulent co-flows and Stokes numbers smaller than about 20, the ratio of the number density of droplets residing within the clusters and voids to the mean number density of the droplets were about 5--6 and 0.7--0.9, respectively. Though increasing the Stokes number to about 140 did not change the ratio of the normalized mean number density of droplets residing within voids, the number density of droplets residing within clusters increased to about 8. The joint PDF of clusters volume normalized by the locally estimated mean value and the mean diameter of the droplets within the clusters showed that although clusters with small volume were occupied by droplets with a broad range of diameters, large-volume clusters accommodated droplets with small variation in their diameter. Overall, the present study extends the findings of past particle-laden flows investigations performed for relatively small Stokes numbers and using 2D measurement techniques to large Stokes numbers relevant to engineering application using 3D diagnostics.

\section*{Declaration of interest}
The authors report no conflict of interest.

\section*{Acknowledgments}
The authors are grateful for the financial support from the University of British Columbia and Canada Foundation for Innovation.

\section*{Appendix A: Estimation of the droplets centre discrepancy for AIPI}\label{Appendix A}

Past investigations that utilized 2D interferometric measurements for droplet positioning, see for example \cite{boddapati2020novel,hardalupas2010simultaneous, rostami2023separate}, have shown that the centres of the imaged circles (corresponding to the droplets) do not match the true centres of the droplets, and as a result, a discrepancy exists in droplet centre identification. In previous investigations, such discrepancy was quantified and the true location of the droplets centres were obtained. Similar to the 2D measurements, here, it was confirmed that the centres of the imaged ellipses (obtained from the AIPI technique) are different from the true location of the droplets centre, with details for obtaining the above centre discrepancies provided in the following.

\begin{figure}
\centerline{\includegraphics[width=1\textwidth]{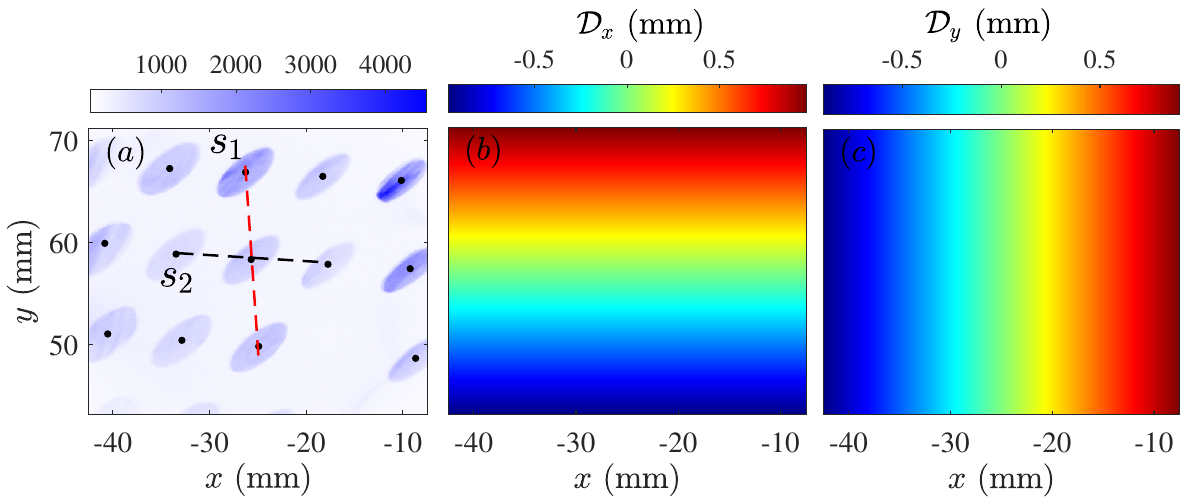}}
\caption{(a) is the image of the target plate captured using the AIPI camera positioned at $z=-20~\mathrm{mm}$. (b) and (c) present the spatial distribution of centre discrepancy along the $x-$ and $y-$axes, respectively.}
\label{Fig:Center discrepency}
\end{figure}

First, a target plate which featured 1~mm diameter holes arranged in a rectangular pattern (spacing of 10~mm) was manufactured and positioned in the $x-y$ plane shown in Fig.~\ref{Fig:Coordinate system}. Then, the camera (that was used for the AIPI technique, i.e. C1 in Fig.~\ref{Fig:Diagnostics}(a)) was positioned at several locations along the $z-$ axis, the plate was shined with an LED backlight, and the images of the target plate were captured, with a sample image corresponding to $z=-20$~mm shown in Fig.~\ref{Fig:Center discrepency}(a). As can be seen, the images of circular holes created on the target plate are tilted ellipses, and these ellipses do not entirely follow the rectangular arrangement of the circular holes. To obtain the mapping relation between the position of the ellipses centres (see the black circular data points in Fig.~\ref{Fig:Center discrepency}(a)) and the centre of the holes on the target plate, the centres of four adjacent ellipses were used to estimate the slope of the connecting lines (see $S_\mathrm{1}$ and $S_\mathrm{2}$ in Fig.~\ref{Fig:Center discrepency}(a)). These slopes were used to obtained the above mapping relation and quantify the centre discrepancies along $x$ and $y$ directions, which are referred to as $\mathcal{D}_x$ and $\mathcal{D}_y$, respectively. The variations of $\mathcal{D}_x$ and $\mathcal{D}_y$ in the $x-y$ plane are shown in Figs.~\ref{Fig:Center discrepency}(b) and (c), respectively. The above experiment was performed for values of $z$ ranging from -25~mm to 25~mm and the values of the centre discrepancies were insensitive to $z$. The results in Figs.~\ref{Fig:Center discrepency}(b) and (c) show that the centre discrepancy is nearly zero at the centre of the region-of-interest and increases linearly towards the edges of this region with the maximum discrepancies being about 1.4~mm near the corners of this region. For each identified ellipse centre, the centre discrepancies in Figs.~\ref{Fig:Center discrepency}(b) and (c) were used to obtain the true location of the droplets centre.

\section*{Appendix B: Calibration of the AIPI for droplet positioning and sizing}
\label{Appendix B}
\begin{figure}
\centerline{\includegraphics[width=1\textwidth]{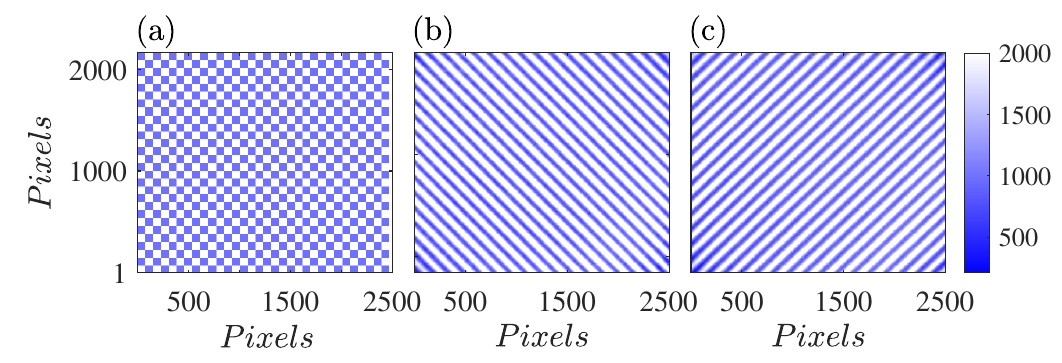}}
\caption{(a) is the checkerboard plate printed and used for the AIPI calibration. (b) and (c) are the images of the checkerboard positioned at focal planes (1) and (2), respectively.} 	
\label{Fig:Calibration1}
\end{figure}

First, the procedure for obtaining the $z$-location of the AIPI focal planes ($c-a$ and $c-b$) as well as the corresponding magnification ratios ($M_1$ and $M_2$) are elaborated. Then, the procedure for measuring the relation between the orientation of the fringe patterns ($\alpha$) and $z$ (see the blue circular data points in Fig.~\ref{Fig:AIPI Data reduction}(e)) is discussed. A checkerboard target plate was prepared, see Fig.~\ref{Fig:Calibration1}(a), printed, and used to obtain the focal planes as well as the magnification ratios. To this end, the target plate was mounted on a translation stage which moved along the $z-$axis. Once the target plate was positioned at focal planes (1) and (2), its corresponding images were slanted lines, which are respectively parallel with the meridian and power axes of the cylindrical lens CL2 in Fig.~\ref{Fig:Diagnostics}(a). The images of the target plate positioned at focal planes (1) and (2) are shown in Fig.~\ref{Fig:Calibration1}(b) and (c), respectively. The distance between the two adjacent slanted lines in Fig.~\ref{Fig:Calibration1}(a) and (b)  were calculated, which were 0.745 and 0.765~mm, respectively. $M_1$ and $M_2$ were obtained by dividing the above-measured distances by the actual distance, which was $\sqrt{2}$~mm. That is $M_1 = 0.745/\sqrt{2} = 0.53$ and $M_2 = 0.765/\sqrt{2} = 0.54$.

\begin{figure}
\centerline{\includegraphics[width=0.6\textwidth]{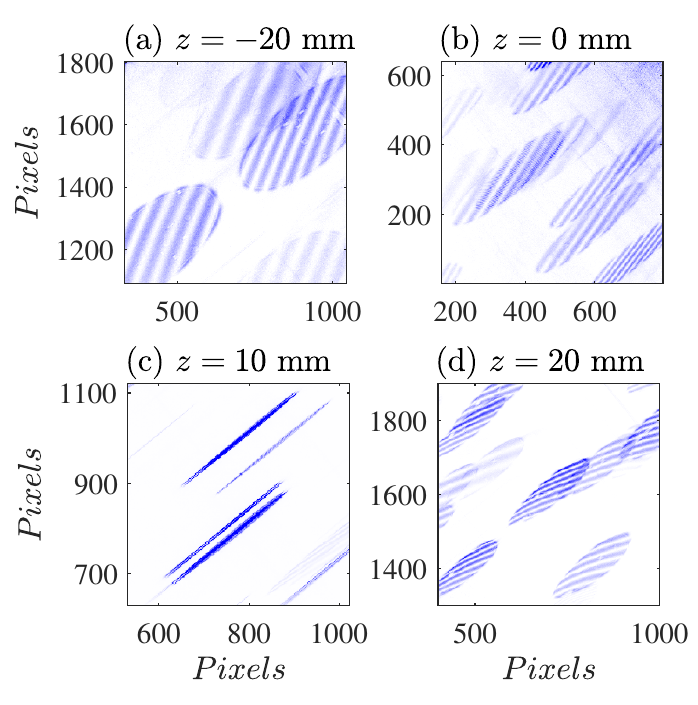}}
\caption{(a)--(d) are the representative cropped images obtained from AIPI for planar laser illumination at $z=$-20, 0, 10, and 20~mm, respectively. The results correspond to mean bulk flow velocity of 9.1~$\mathrm{ms^{-1}}$.}
\label{Fig:Calibration2}
\end{figure}

In order to assess the analytically obtained calibration relation in Eq.~(\ref{Eq:zlocation}), which relates the depth of the droplets and the angle of the fringes, separate calibration experiments were performed. Specifically, the spherical lens (SL3) in Fig.~\ref{Fig:Diagnostics} was removed and the AIPI was performed for a planar illumination of the spray. This was conducted for several planes along the $z-$ axis by positioning the camera at several locations along this axis, with the sample collected fringe patterns shown in Fig.~\ref{Fig:Calibration2}. The results in (a)--(d) are the cropped AIPI images and correspond to $z$= -20, 0, 10, and 20~mm, respectively. As can be seen, for a fixed value of $z$, the angles of the fringe patterns within an image do not change, since the illumination was performed at a plane. For each examined value of $z$, 500 images were collected, the angles for all imaged fringe patterns were obtained, and the variation of the mean angle versus $z$ was presented in Fig.~\ref{Fig:AIPI Data reduction}(e), with the largest error bar being twice the largest standard deviation of $\alpha$ obtained from the calibration measurements. This was performed for all tested mean bulk flow velocities, and it was obtained that the relation between the $z$ location of the droplets versus $\alpha$ is independent of the tested mean bulk flow velocities.

\section*{Appendix C: Uncertainty in estimating the droplet location and its diameter}\label{Appendix C}

\begin{figure}
\centerline{\includegraphics[width=0.9\textwidth]{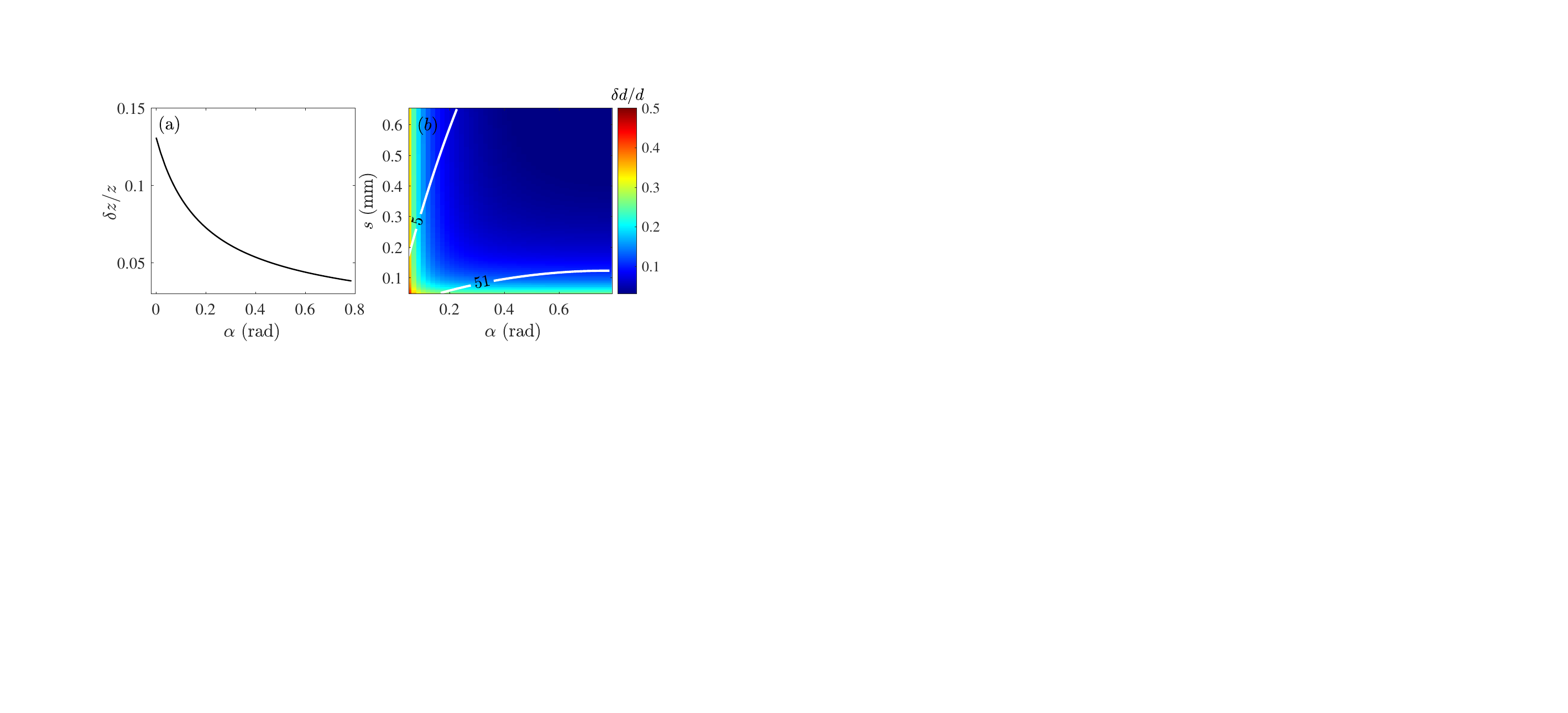}}
\caption{(a) is the variation of relative uncertainty in measuring the droplet $z-$location versus the fringe angle. (b) presents the variation of relative uncertainty in estimating the droplet diameter versus the fringe angle and fringe spatial spacing. Overlaid on (b) are the contours of the minimum measurable and most probable droplet diameter, see the white colour curves.}
\label{Fig:Uncertanity}
\end{figure}

The uncertainty in calculating the droplet position along the $z-$axis was estimated following~\cite{coleman2018experimentation}, assuming $\alpha$, $c-b$, and $c-a$ are independent measurable variables. Specifically, the uncertainty in calculating the droplet location along the $z-$axis was calculated from
 \begin{equation}
  \label{Eq:dz}
 \delta z= \sqrt{\left[\frac{\partial z}{\partial \alpha}\delta \alpha\right]^{2}+\left[\frac{\partial z}{\partial (c-b)}\delta (c-b)\right]^{2}+\left[\frac{\partial z}{\partial (c-a)}\delta (c-a)\right]^{2}},
\end{equation}
with terms ${\partial z}/{\partial \alpha}$, ${\partial z}/{\partial (c-b)}$, and ${\partial z}/{\partial (c-a)}$ being the sensitivities in measuring the $z$ location of the droplets with respect to the independent variables. $\delta \alpha$ is the uncertainty in measuring the fringe orientation and was estimated to be $1^{\mathrm{o}}$ using the imaging resolution. $\delta (c-b)$, and $\delta (c-a)$ are the uncertainties in measuring $c-b$ and $c-a$ which were about 1~mm. The variation of relative uncertainty in measuring the droplet location along the $z-$axis ($\delta z/z$) versus the fringe orientation was estimated using Eq.~(\ref{Eq:dz}), with the corresponding results presented in Fig.~\ref{Fig:Uncertanity}(a). As can be seen, the largest relative uncertainty is about 13\% and corresponds to droplets with $\alpha=0$ (i.e. droplets that are located at focal plane $z=10$~mm). 

Following \cite{coleman2018experimentation}, Eqs.~(\ref{Eq:d})~and~(\ref{Eq:Fz}) were used to estimate the uncertainty in measuring the droplet diameter ($\delta d$) from AIPI which is given by
 \begin{equation}
  \label{Eq:d2}
 \delta d= \sqrt{\left[\frac{\partial d}{\partial \alpha}\delta \alpha\right]^{2}+\left[\frac{\partial d}{\partial s}\delta s\right]^{2}+\left[\frac{\partial d}{\partial \theta}\delta \theta\right]^{2}}.
\end{equation}
In Eq.~(\ref{Eq:d2}), the terms ${\partial d}/{\partial \alpha}$, ${\partial d}/{\partial s}$, and ${\partial d}/{\partial \theta}$ are the sensitivities of the droplet diameter with respect to the independently measured variables. $\delta \alpha = 1^\mathrm{o}$, $\delta s=13~\mu m$, and $\delta \theta = 1^\mathrm{o}$ are the uncertainties in measuring $\alpha$, $s$, and $\theta$, respectively. The relative uncertainty in measuring the droplet diameter ($\delta d/d$) can be estimated using Eq.~(\ref{Eq:d2}) and is given by 
\begin{equation}
	\label{Eq:duncertanity}
	\frac{\delta d}{d}= \sqrt{\left[\frac{dF}{d\alpha}\frac{1}{F}\delta \alpha\right]^{2}+\left[\frac{1}{s}\delta s\right]^{2}+\left[\frac{dg}{d \theta}\delta \theta\right]^{2}},
\end{equation}
with the formulation for $F(\alpha)$ obtained by combining Eq.~(\ref{Eq:Fz})~and~(\ref{Eq:zlocation}). In Eq.~(\ref{Eq:duncertanity}), $g = 2\lambda F/(sd)$. In the present study, $\theta$ was unchanged and equals $70^\mathrm{o}$. The relative uncertainty versus $\alpha$ and $s$ is presented in Fig.~\ref{Fig:Uncertanity}(b). The contours of minimum measurable diameter from AIPI ($d_\mathrm{min}=5~\mu \mathrm{m}$) and the most probable droplet diameter ($d=51~\mu \mathrm{m}$) are also overlaid on Fig.~\ref{Fig:Uncertanity}(b) using the white colour curves. The results in Fig.~\ref{Fig:Uncertanity}(b) show that the maximum relative uncertainty in reporting the most probable droplet diameter is about 27\% .  

\bibliography{Bib}
\bibliographystyle{jfm}
\end{document}